\documentclass[12pt,fleqn]{article}
\usepackage{amssymb}
\usepackage{graphicx}
\usepackage{amsmath}
\usepackage{amsfonts}

\begin{document}

\begin{centering}
{\leftskip=2in \rightskip=2in
{\large \bf Introduction to Quantum-Gravity Phenomenology}}\\
\bigskip
\bigskip
\bigskip
\medskip
{\small {\bf Giovanni AMELINO-CAMELIA}}\\

\bigskip
{\it Dipart.~Fisica Univ.~La Sapienza and Sez.~Roma1 INFN}\\
{\it P.le Moro 2, I-00185 Roma, Italy}

\end{centering}

\vspace{0.7cm}

\begin{center}
\textbf{ABSTRACT}
\end{center}

\baselineskip 11pt plus .5pt minus .5pt

{\leftskip=0.6in \rightskip=0.6in {\footnotesize
After a brief review of the first phase of
development of Quantum-Gravity Phenomenology,
I argue that this research line
is now ready to enter a more advanced phase: while at first it was
legitimate to resort to heuristic order-of-magnitude
estimates, which were sufficient to establish that
sensitivity to Planck-scale effects can be achieved,
we should now rely on detailed analyses
of some reference test theories.
I illustrate this point in the specific example of
studies of Planck-scale modifications of the energy/momentum
dispersion relation, for which I consider two test theories.
Both the photon-stability analyses and the Crab-nebula
synchrotron-radiation analyses, which had raised high hopes
of ``beyond-Plankian'' experimental bounds,
turn out to be rather ineffective in constraining
the two test theories.
Examples of analyses which can provide constraints of
rather wide applicability are
the so-called ``time-of-flight analyses'', in the context of
observations of gamma-ray bursts,
and the analyses of the cosmic-ray spectrum near the
GZK scale.
}}

\vspace{24pt}
\newpage

\baselineskip 12pt plus .5pt minus .5pt
\pagenumbering{arabic}

\setcounter{footnote}{0} \renewcommand{\thefootnote}{\alph{footnote}}

\pagestyle{plain}

\section{From ``Quantum Gravity beauty contests''
to Quantum Gravity Phenomenology}
The ``quantum-gravity problem" has been studied
for more than 70 years~\cite{stachelearly}
assuming that no guidance could
be obtained from experiments. This in turn led to
the assumption that the most promising
path toward the solution of the problem would be the
construction and analysis of very ambitious theories
(some would call them ``theories of everything"),
capable of solving at once
all of the issues raised by the coexistence of gravity
and quantum mechanics.
In other research areas the availability
of experimental data challenging the current theories
encourages theorists to
propose phenomenological models which solve the
experimental puzzles, even when some aspects of the models are
not fully satisfactory from a conceptual perspective.
Often those apparently unsatisfactory models
turn out to provide an important starting point
for the identification of the correct (and conceptually satisfactory)
theoretical description of the new phenomena.
But in this quantum-gravity research area, since there was
no experimental guidance, it was inevitable for theorists
to be tempted into trying to identify the correct theoretical
framework relying exclusively on some criteria
of conceptual compellingness.
Of course, tempting as it may seem, this strategy would not be
acceptable for a scientific endeavor. Even the most
compelling and conceptually satisfying theory could not
be adopted without experimental confirmation.

The mirage (occasionally mentioned at relevant seminars)
that one day within an ambitious
quantum-gravity theory one
might derive from first principles a falsifiable prediction for
the mundane realm of doable experiments could give some ``scientific
legitimacy" to
these research programmes, but this possibility
never materialized. It may indeed be just a mirage.
There are several occasions when a debate between advocates of
different ambitious
quantum-gravity theories shapes up in a way similar to the discussion
between advocates of different religions. And often in the media
the different approaches are compared on the basis of the ``support''
they have in the community: one says ``the most popular approach
to the quantum gravity problem'' rather than ``the approach that
has had better success reproducing experimental results''.
So, it would seem, the Quantum Gravity problem is to be solved
by an election, by a beauty contest, by a leap of faith.

Over the last few years a growing
number of research groups have attempted to tackle the
quantum-gravity problem with an approach which is more
consistent with the traditional strategy of scientific work.
Simple (in some cases even simple-minded)
non-classical pictures of spacetime are being analyzed
with strong emphasis on their observable predictions.
Certain classes of experiments have been shown to
have extremely high sensitivity to
some non-classical features of spacetime.
We now even have (see later)
some first examples of experimental puzzles
whose solution is being sought also within simple ideas involving
non-classical pictures of spacetime.
The hope is that by trial and error, both on the theory side
and on the experiment side, we might eventually stumble upon
the first few definite (experimental!) hints on the quantum-gravity
problem.

Quantum gravity phenomenology
requires of course a combination of theory and experiments.
It does not adopt any
particular prejudice concerning the structure of spacetime
at short distances (in particular, ``string
theory"~\cite{string1,string2}, ``loop quantum
gravity"~\cite{crLIVING,ashtNEW,leeLQGrev,thieREV}
and ``noncommutative geometry"~\cite{connesbook,majidbok}
are seen as equally deserving
mathematical-physics programmes),
but of course must follow as closely as possible the
few indications
that these ambitious quantum-gravity theories provide.
One here is guided by the expectation that quantum-gravity research should
proceed just in the old-fashioned way
of scientific work: through small incremental
steps starting from what we know and combining mathematical-physics
studies with experimental studies to reach deeper and deeper layers
of understanding of the problem at hand (in this case
the short-distance structure of spacetime and the laws that govern it).

The most popular quantum-gravity approaches, such as
string theory and loop quantum gravity, could be described as ``top-to-bottom
approaches" since they start off with some key assumption about
the structure of spacetime at scales that
are some 17 orders of magnitude
beyond the scales presently accessible experimentally, and then they should
work their way back to the realm of doable experiments.
With ``quantum gravity phenomenology" I would like to refer to all
studies that are intended to contribute
to a ``bottom-to-top approach"
to the quantum-gravity problem.
Since the problem at hand is really difficult (arguably the most challenging
problem ever faced by the physics community)
it appears likely that the two complementary
approaches might combine in a useful way: for the ``bottom-to-top approach"
it is important to get some guidance from the (however tentative)
indications emerging
from the ``top-to-bottom approaches", while for ``top-to-bottom approaches"
it might be useful to be alerted by quantum-gravity phenomenologists
with respect to the type of new effects that could be most stringently
tested experimentally (it is hard for ``top-to-bottom approaches" to
obtain a complete description of ``real" physics, but perhaps it
would be possible to dig out predictions on some specific spacetime features
that appear to deserve special attention in light of the corresponding
experimental sensitivities).

In these lectures I give a ``selected-topics'' introduction
to this ``Quantum Gravity Phenomenology".
I will in particular stress that, while the first few years
of work in this area, the ``dawn'' of quantum-gravity
phenomenology\cite{polonpap},
were necessarily based on rather preliminary analyses,
with the only objective of establishing the point that
Planck-scale sensitivity could be achieved in some cases,
we should now gear up for a more ``mature'' phase of work
on quantum-gravity phenomenology, in which the development
and analysis of some carefully crafted test theories takes center stage.

\section{Quantum Gravity Phenomenology}
In this section I describe the key objectives of quantum-gravity
phenomenology and sketch out its strategy in the search of
the first manifestation of a quantum property of spacetime.
I also start introducing my argument that we should now
move from the ``dawn'' of quantum-gravity phenomenology
to a more ``mature'' quantum-gravity phenomenology,
in which a key role is played by
the development
and analysis of some carefully crafted test theories.

\subsection{Planck-scale quantum properties of spacetime}
The first step for the identification of experiments
relevant for quantum gravity is of course the identification
of the characteristic scale of this new physics.
This is a point on which we have relatively robust
guidance from theories and theoretical arguments:
the characteristic scale at which non-classical properties
of spacetime physics become large (as large as the classical properties
they compete with) should be
the Planck length $L_p \sim 10^{-35}m$ (or equivalently
its inverse, the Planck scale $E_p \sim 10^{28} eV$).
The key challenge for quantum-gravity phenomenology must be
the one of establishing ways to provide sensitivity to
Planck-scale non-classical
properties of spacetime.

I will call ``quantum'' properties of spacetime all effects
which represent departures from a classical picture of spacetime.
This is after all what is commonly done in the literature,
where authors often use the name ``quantum properties of spacetime''
because of the expectation that some of the familiar features
of quantization, which showed up everywhere else in physics, should
eventually also play a role in the description of spacetime.
There is no guarantee that the non-classical properties
of spacetime will take the shape of some sort of proper spacetime
quantization. But, as long as this is
understood, the use of the spacetime-quantization terminology
does no arm.

Of course, the search of a solution of the quantum-gravity problem
can benefit also from other types of experimental insight,
and therefore the scopes of quantum-gravity phenomenology
must go even beyond its key quantum-spacetime challenge.
In particular, quantum gravity should also provide a consistent
description of the quantum properties of particles in presence
of strong (or anyway nonnegligible) classical gravity fields.
This type of context at
the ``Interface of Quantum and [classical] Gravitational Realms"~\cite{iqgr}
has been the subject of a rather sizeable literature for several
decades.
When quantum properties of spacetime are not relevant
for the analysis the insight one can gain for the
quantum-gravity problem is of more limited impact, but it is
of course still valuable.
Indeed a valuable debate on the fate of the Equivalence Principle
in quantum gravity was ignited already in the mid 1970s with the renowned
experiment performed by Colella, Overhauser and Werner~\cite{cow}.
That experiment has been followed by several modifications
and refinements (often labeled ``COW experiments'' from the initials
of the scientists involved in the first experiment)
all probing the same basic physics, {\it i.e.} the validity
of the Schr\"{o}dinger equation
\begin{eqnarray}
\left[ - \left( {\hbar^2 \over 2 \, M_I} \right) \vec{\nabla}^2
+ M_G \, \phi(\vec{r}) \right] \psi(t,\vec{r}) = i \, \hbar \,
{\partial \, \psi(t,\vec{r}) \over \partial t}
\label{coweq}
\end{eqnarray}
for the description of the dynamics of matter (with wave
function $\psi(t,\vec{r})$) in presence of the Earth's
gravitational potential $\phi(\vec{r})$.
[In (\ref{coweq}) $M_I$ and $M_G$ denote the inertial
and gravitational mass respectively.]

The COW experiments exploit the fact that
the Earth's gravitational potential puts together the contributions
of a very large number of particles
and as a result, in spite of its per-particle weakness,
the overall gravitational field is large
enough\footnote{Actually the effect turns out to be observably large
because of a double ``amplification": the first, and most significant,
amplification is the mentioned coherent addition of gravitational
fields generated by the particles that compose the Earth,
the second amplification~\cite{dharamCOW1} involves the ratio between
the wavelength of the particles used in the COW experiments
and some larger length scales involved in the experimental
setup.}
to introduce observable effects.
The relevance of these experiments for the debate
on the Equivalence Principle will not be discussed here,
but has been discussed in detail by several authors
(see, {\it e.g.}, Refs.~\cite{sakurai,gasperiniEP,dharamEP}).
I here just bring to the reader's attention
a recent experiment which appears to indicate a violation
of the Equivalence Principle~\cite{cowEPviol}
(but the reliability of this experimental result is still
being debated),
and some ideas for intruiging
new experiments~\cite{dharamCOW1,dharamCOW2}
of the COW type.
I should also mention for completeness
the related work on
the interplay
between classical general relativity and quantum mechanics of
nongravitational degrees of freedom
reported in Refs.~\cite{anan1,anan2}.

Another possibility that, even though it is in contrast with
the idea of Planck-scale quantum properties of spacetime,
deserves some exploratory effort by those working in
quantum-gravity phenomenology is the one of scenarios
in which the standard
estimate of the quantum-gravity scale as the Planck scale
turns out to be too pessimistic.
There is (at present) no compelling argument in support
of the idea that
the quantum-gravity scale should be effectively lowered,
but this possibility cannot be excluded.  In particular,
some recent studies~\cite{led}
found a mechanism that would allow to lower significantly
the quantum-gravity energy scale, several orders of magnitude
below the Planck energy scale.
This mechanism relies of the hypothesis of ``large extra dimensions''
which is not in any way ``natural" (not even in the eyes of
the scientists who proposed it), but it can be used to
provide an example of a workable scenario for a low
scale of quantum-gravity effects.

For the rest of these lectures I will however focus on what
I described as the key challenge: the search of
Planck-scale quantum properties of spacetime.

\subsection{Identification of experiments}
\noindent
Unfortunately, in spite of more than 70 years of theory
work on the quantum-gravity problem, and a certain proliferation
of theoretical frameworks being considered, there is only
a small number of physical effects that have been considered
in the quantum-gravity literature. Moreover, most of these effects
concern strong-gravity contexts, such as black-hole physics
and big-bang physics, which are exciting at the level
of conceptual analysis and development of formalism,
but are not very promising
for the actual (experimental) discovery of manifestations of
non-classical properties of spacetime.

While it is likely that the largest quantum-gravity effects
should be present in large-curvature situations, it only takes
a little reasoning to realize that
we should give priority to quantum-gravity effects that
modify our description of (quasi-)Minkowski spacetime.
The effects will perhaps
be smaller than, say, in black hole physics (in some
aspects of black hole physics quantum-gravity effects
might be as large as classical physics effects), but we are
likely to be better off considering quasi-Minkowski spacetimes,
for which the quality of the data we can obtain is extremely high.

In the analysis of flat-spacetime processes,
involving particles with energies that are inevitably much
lower than the Planck energy scale,
we will have to deal with a large suppression of
quantum-gravity effects,
a suppression which is
likely to take the form of some power of the ratio
between the Planck length and the wavelength of the
particles involved.
The presence of these suppression factors
on the one hand reduces sharply our chances of finding
quantum-gravity effects, but on the other hand simplifies
the problem of identifying promising experimental contexts,
since these experimental contexts must enjoy very
special properties which would not go easily unnoticed.
For laboratory experiments
even an optimistic estimate of these suppression factors
leads to a suppression of order $10^{-16}$, which one obtains
by assuming (probably already using some optimism)
that at least some quantum-gravity effects are only linearly
suppressed by the Planck length, and taking as particle wavelength
the shorter wavelengths we are able to produce ($\sim 10^{-19}m$).
In astrophysics (which however limits one to ``observations"
rather than ``experiments") particles of shorter wavelength
are being studied, but even for the highest energy cosmic rays,
with energy of $\sim 10^{20}eV$ and therefore wavelengths
of $\sim 10^{-27}m$, a suppression of the type $L_p/\lambda$
would take values of order $10^{-8}$.
It is mostly as a result of this type of considerations
that traditional quantum-gravity reviews considered
the possibility of experimental studies
of Planck-scale effects with unmitigated
pessimism~\cite{chrisreview}.

However, the presence of large suppression factors
surely cannot suffice for drawing any conclusions.
Even just looking within the subject of particle physics
we know that certain types of small effects can be studied,
as illustrated by the example of the remarkable limits
obtained on proton instability. Outside of fundamental physics more
success stories of this type are easily found. Think for example
of brownian motion, where some unobservably small micro-processes
lead to an effect which is observable on macroscopic scales.

It is hard but clearly not impossible to find
experimental contexts in which there is effectively an
amplification of the small effect one intends to study.
The prediction of proton decay within certain grandunified theories
of particle physics is really a small effect, suppressed by the fourth
power of the ratio between the mass of the proton and
grandunification scale, which is only
three orders of magnitude smaller than the Planck scale.
In spite of this horrifying suppression,
of order $[m_{proton}/E_{gut}]^4 \sim 10^{-64}$,
with a simple idea we have managed to acquire full sensitivity
to the new effect: the proton lifetime predicted by grandunified
theories is of order $10^{39}s$ and ``quite a few" generations
of physicists should invest their entire lifetimes staring at a single
proton before its decay, but by managing to keep under observation
a large number of protons (think for example of a situation in which
$10^{33}$ protons are monitored)
our sensitivity to proton decay is dramatically increased.
In that context the number of protons is the
dimensionless quantity that works as ``amplifier" of the new-physics
effect. Similar considerations explain the success of brownian-motion
studies already a century ago.

We should therefore focus our attention\cite{polonpap}
on experiments which have something to do with spacetime
structure
and that host an ordinary-physics dimensionless quantity
large enough that (if we are ``lucky") it could amplify
the extremely small effects we are hoping to discover.
So there is clearly a first level of analysis
in which one identifies experiments with this rare
quality, and a second level of analysis in which one
tries to establish whether indeed the candidate ``amplifier"
could possibly amplify effects connected with spacetime
structure.

\subsection{Prehistory of quantum gravity phenomenology}
Clearly a good phenomenological programme
must be able to falsify theories.
Although it is already noteworthy that some candidate
quantum-gravity effects could at all be looked for in the data,
this would not be so significant if we were not able to use
these data to constrain the work of theorists, to falsify some
theoretical pictures.
The fact that, toward the end of the 1990s,
it was convincingly argued that this could be done
brought the idea of ``Planck-scale tests''
to center stage in quantum-gravity research.
The fact that we could plausibly gain insight on
Planck-scale physics is now widely acknowledged in the quantum-gravity
community.
Up to 1997 or 1998 there had already been some works
on the possibility to find experimental evidence of
some Planck-scale effects, but the relevant data analyses
did not in reverse have the capability to falsify
any quantum-gravity picture
and the relevant research remained at
the margins of the mainstream quantum-gravity literature.

A first example of these works of the ``prehistory of quantum-gravity
phenomenology'' is provided by a certain type of
investigation of Planck-scale departures from
CPT symmetry using
the neutral-kaon and the neutral-B
systems~\cite{ehns,huetpesk,emln,floreacpt}.
These pioneering works were based on the realization
that in the relevant neutral-meson systems
a Planck-scale departure from CPT symmetry could in principle
be amplified; in particular, the neutral-kaon system
hosts the peculiarly small mass difference between
long-lived and the short-lived
kaons $|M_L - M_S|/M_{L,S} \sim 7 {\cdot} 10^{-15}$.
The quantum-gravity picture usually advocated in these studies
is the one of a variant of the string-theory picture,
which relies on noncritical strings, in the so-called ``Liouville''
approach~\cite{emln,emn}.
This is an ambitious attempt for a theory of everything, which,
while based on an appealing view of the quantum-gravity problem,
is for the most part untreatable, at least with current techniques.
The departures from CPT symmetry cannot be derived from the theory,
but one can provide tentative evidence that the structure of the theory
should accommodate such departures.
As a result one is forced to set up a multi-parameter phenomenology
which looks for the new effects, but a negative outcome of the
experiments could not be used to falsify the framework which
is at the root of the analysis.

Similar remarks apply to the other pioneering
studies reported in Refs.~\cite{kostcpt},
which find their original motivation in
some aspects of ``string field theory''~\cite{kostcpt}.
Also the string-field-theory formalism is very ambitious
and too hard to handle. A multi-parameter phenomenology is
necessarily set up~\cite{kostcpt},
and a negative outcome of the
experiments could not be used to falsify the framework which
is at the root of the analysis.
Besides the falsifiability issue, this phenomenology may not appeal
to many quantum-gravity researchers because it mainly focuses
on the ``Standard Model Extension'',
whose key assumption~\cite{kostrenorm}
is the renormalizability of the underlying field theory.
The assumption of renormalizability limits one to effects
that area described in terms of operators of dimension 4 and lower,
whereas most quantum-gravity researchers expect Planck-scale-suppressed
effects described in terms of operators of dimension 5 and higher.

A third equally-deserving entry in my list
of pioneers of the ``prehistory of quantum-gravity
phenomenology'' is the work reported in Refs.~\cite{perci,percistru}
which explored the general issue of how certain effectively
stochastic properties of spacetime would affect
the evolution of quantum-mechanical states.
The guiding idea was that stochastic processes could provide
an effective description of quantum spacetime processes.
The implications of these stochastic properties
for the evolution of quantum-mechanical states
were modelled in Refs.~\cite{perci,percistru}
via the formalism of ``primary  state diffusion'',
but only rather crude models turned out to be treatable.
As also emphasized by the authors, the
crudeness of the models is such that all conclusions are to be
considered as tentative at best, and this is one more instance
in which a negative outcome of the
experiments could not be used to falsify the framework which
is at the root of the analysis.

The three research lines I discussed in this subsection as
examples of ``prehistory of quantum-gravity
phenomenology'' showed convincingly that the possibility of
stumbling upon an experimental manifestation of Planck-scale
effects could not be excluded.
On the other hand they proved to be insufficient for the
birth a genuine, fully articulated, phenomenological programme.
In that regard their key common limitation was the mentioned
fact that it appeared that the relevant experiments could
not falsify the relevant theories.
Moreover, it often appeared that these studies were establishing
that some not-much-studied quantum-gravity approaches could
lead to observable effects, as a way to distinguish them from
the most popular quantum-gravity ideas which would remain
untestable. Of course, the interest of the community grew
when it became apparent that a rather large variety of
quantum-gravity ideas could lead to observable effects
(and could be falsified). A sizeable community now
works under the assumption that the presence of observably-large
quantum-gravity effects is not a peculiar feature of some
out-of-mainstream quantum-gravity approaches: it is a property of
most quantum-gravity approaches, including some of the
most popular ones.

\subsection{The dawn of quantum gravity phenomenology}
\noindent
The research lines discussed in the previous subsections
had been establishing that it was not inconceivable to
use data within our reach (inevitably involving particles
with energies much lower than the Planck energy scale)
to find evidence of a Planck-scale effect.
However, while these research did ignite a lively interest by some
experimentalists (see, {\it e.g.}, Refs.~\cite{web1,web2}),
they went largely unnoticed by mainstream quantum-gravity research.
As stressed above, this was likely due to the fact that
they were incomplete proposals from the viewpoint of phenomenology,
because the test theories could not be really falsified,
and because the relevant test theories were all outside mainstream
quantum-gravity research, so that the fact that Planck scale effects
could be seen appeared to be a peculiar property of out-of-mainstream
theories.
On the other hand,
clearly those research lines were starting set the stage
for a wider and more developed phenomenological effort,
which indeed came to existence toward the end of the 1990s.
When, indeed starting toward the end of the 1990s,
the case for falsifiability of some Planck-scale
models started to be built, and first evidence of testability of
mainstream quantum-gravity proposals emerged,
a corresponding quick growth of interest emerged in the community.
This is perhaps best illustrated by comparing
authoritative quantum-gravity reviews published up to the
mid 1990s (see, {\it e.g.}, Ref.~\cite{chrisreview})
and the corresponding reviews published over the last couple of
years~\cite{ashtereview,crHISTO,leePW,carlip}.

This ``dawn'' of quantum-gravity phenomenology has revolved around
a growing number of experimental contexts in which Planck-scale
effects are being sought.
Among the most popular such proposals let me mention, as a few
noteworthy examples, the
studies of in-vacuo dispersion using gamma-ray
astrophysics~\cite{grbgac,billetal},
studies of laser-interferometric \index{interferometer}
limits on quantum-gravity
effects~\cite{gacgwi,bignapap,nggwi,bignapatwo,gaclaem},
studies of the role of quantum-gravity effects in the determination
of the energy-momentum-conservation threshold conditions
for certain particle-physics processes~\cite{kifu,ita,aus,gactp,jaco},
and studies of the role of quantum gravity in the determination
of particle-decay amplitudes~\cite{gacpion,orfeupion}.

The idea of looking for Planck-scale departures from CPT symmetry
continues to be pursued, but in that context we are still lacking
an analysis showing how a quantum-gravity model could be falsified
on the basis of such CPT studies. This is essentially due to
some technical challenges in establishing what exactly
happens to CPT symmetry within a given Planck-scale picture.
It is often easy
to see that CPT is affected, but one is then unable to establish
how it is affected.

As I shall stress again later, among all these research lines
a special role in the development of quantum-gravity phenomenology
is being played by studies of
the role of quantum-gravity effects in the determination
of the energy-momentum-conservation threshold conditions
for certain particle-physics processes.
In fact, in these studies we have stumbled upon
a first example of experimental puzzle
whose solution could plausibly be sought within quantum-gravity
phenomenology. This of course marked an important milestone
for quantum-gravity phenomenology. The relevant context
is the one of the process of
photopion production, $p + \gamma \rightarrow p + \pi^0$,
which, as discussed later in these lectures,
plays a crucial role in the analysis of the cosmic-ray
spectrum.
An apparent ``anomaly'' in the observed cosmic-ray spectrum
could be naturally described in terms of Planck-scale effects.
Of course, it is not unlikely that
this ``anomaly'' might fade away, as better
data on cosmic rays become available,
but it is nonetheless an important sign of maturity for quantum-gravity
phenomenology that some data invite interpretation as a possible
manifestation of Planck-scale physics. Chances are the first few
such ``candidate anomalies'' will turn out to be incorrect, but eventually
one lucky instance could be encountered.

\subsection{The maturity of
Quantum Gravity Phenomenology: test theories}
\noindent
The fact that quantum-gravity phenomenology is already being
considered in attempts to solve present experimental puzzles
is indeed a clear indication of progress toward the maturity of
the field, but in many respects the field is still rather immature.
The first challenge for quantum-gravity phenomenologists
was to establish convincingly that there is a chance to test
Planck-scale effects, and this type of argument can legitimately be
based on intuitive order-of-magnitude analyses.
However, at this point a rather large community
acknowledges that quantum-gravity phenomenology has a chance,
so the first challenge was successfully overcome,
and we must now shift gear. There is very little more to be
gained through rudimental back-of-the-envelope analyses.
The standards of quantum-gravity phenomenology
must be raised to the ones adopted in other branches of phenomenology,
such as particle-physics phenomenology.

In these lectures I shall in particular emphasize the importance
of adopting some reference test theories. If an effect is described
only vaguely, without the support of an associated test theory,
then the experimental limits that can be claimed
are of correspondingly uncertain significance.
As it has happened in the recent quantum-gravity-phenomenology literature,
different authors may end up claiming
different limits on ``the same effect'' simply because they are actually
adopting different test theories and therefore they are truly
analyzing different effects.
This type of phenomenology clearly would not help us gain any
insight on Planck-scale physics. The main task of phenomenology
is to provide to the theorists working at the development
of the theories information
on what is and what is not consistent with experimental data.
Phenomenology essentially provides some boundaries within which
formal theorists are then forced to work. A theory which would
predict effects inconsistent with some data is abandoned.
But if this boundaries are not clearly drawn, if the experimental limits
are placed on ``effects'' which are not rigorously defined within
the context of a test theory, then they are correspondingly useless
for the development of theories.

The discussion here reported in Section~4 will illustrate this point
in a specific context.

\section{Some candidate Quantum-Gravity effects}
Before focusing, in the next section, on an
example of ``quantum-gravity-phenomenology exercise'',
it seems appropriate to list at least a few of the candidate
quantum-gravity effects that find motivation in the
literature.

Testing these effects will be the main task of quantum-gravity
phenomenology. While here I will discuss these effects
at a rather rudimentary and intuitive level, so that my remarks would
apply to a variety of approaches to the quantum-gravity problem,
clearly in each theory these effects may take a different form,
and in setting up a phenomenology for these effects it will be crucial
to develop some corresponding test theories.

In providing motivation for the study of these effects I could use
a large variety of arguments; however, I find preferable to
show that these effects can be motivated already on the basis
of the most plausible of all hypotheses concerning the quantum-gravity
problem: the hypothesis that some of the incarnations of the ``quantum''
idea (such as discretization and noncommutativity of observables)
should find place also in the description of spacetime.

\subsection{Planck-scale departures from Lorentz symmetry}
Perhaps the most debated possibility for a quantum spacetime,
possibly intended as Planck-scale discrete or
Planck-scale noncommutative spacetime,
is the one of Planck-scale departures from Lorentz symmetry.

The continuous symmetries of a spacetime reflect of course
the structure of that spacetime. Ordinary Lorentz symmetry
is governed by the single scale that sets the structure
of classical Minkowski spacetime, the speed-of-light
scale $c$. If one introduces additional structure
in a flat spacetime its symmetries will be accordingly
affected. This is particularly clear
for some simple ideas concerning a Planck-scale discretization of
spacetime~\cite{hooftlorentz}. Continuous symmetry transformations
are clearly at odd with a discrete network of points.

For different reasons,  Lorentz symmetry is also often at odds
with spacetime noncommutativity. In particular,
it appears that in certain cases
the noncommutativity length scale~\cite{gacdsr}
(possibly the Planck scale), in addition to $c$,
affects the laws of transformation between inertial observers,
and infinitesimal
symmetry transformations are actually described in terms
of the new language of Hopf algebras~\cite{majrue,kpoinap},
rather than by the Poincar\'{e} Lie algebra.
The type of spacetime quantization provided by noncommutativity
may therefore lead to a corresponding ``symmetry
quantization'': the concept of Lie-algebra symmetry is in fact
replaced by the one of Hopf-algebra symmetry.

In a large number of recent studies of noncommutative
spacetimes it has indeed been found that the Lie-algebra
Poincar\'{e} symmetries are either broken to a smaller symmetry
Lie algebra or deformed into Hopf-algebra symmetries.

For what concerns the idea of spacetime
discretization the most developed quantum-gravity picture
is the one of Loop Quantum Gravity, which
does not predict a rigid discrete network of spacetime points,
but introduces discretization in a more sophisticated
way: the spectra of areas and volumes are discretized,
while spacetime points
loose all possible forms of identity.
It appears that even this more advanced form of discretization
is incompatible with classical Lorentz symmetry;
in fact, a growing number of
loop-quantum-gravity studies has been reporting~\cite{gampul,mexweave}
evidence of Planck-scale departures from Lorentz symmetry (although
the issue remains subject to further scrutiny).

\subsection{Planck-scale departures from CPT symmetry}
The fact that our low-energy\footnote{Since it is often obvious
from the context, I will sometimes avoid
specifying ``energies that are low with respect to the Planck
energy scale'' and simply
write ``low-energy''. With ``high-energy particles''
instead I will not mean ``particles with energy higher
than the Planck energy scale'' (a situation which we never encounter),
but rather the case of particles with energy rather close to (but
still lower than) the Planck scale.}
observations are consistent with CPT
symmetry is not a miracle: as codified by the
CPT theorem, a Lorentz-invariant local quantum field theory
is inevitably CPT invariant.
The fact that quantum gravity, the ``unification'' of gravity and
quantum theory, invites us to consider Planck-scale
departures from Lorentz symmetry (as stressed above)
and Planck-scale departures from locality (as natural in a
discrete-spacetime theory)
opens the door for Planck-scale departures from CPT
symmetry.

While this general argument is rather robust, it is not always easy
to establish what is the fate of CPT symmetry in a given
quantum-gravity approach.
For example in Loop Quantum Gravity
the analysis of (the various alternative ideas) on
coupling ordinary particles to gravity has not yet advanced
to the point of allowing a robust description
of $C$ transformations.
On the other hand there are examples in which some progress
in the analysis of CPT transformations has been achieved
and evidence of departures from CPT symmetry is found.
This is for example the case \index{$\kappa$-Minkowski space-time}
of $\kappa$-Minkowski noncommutative
spacetime, where one can clearly see~\cite{gacmajid} a modification
of P transformations.

Since I am not considering CPT symmetry \index{CPT symmetry}
in the remainder of
these lectures let me mention here that, besides the
neutral-kaon and neutral-B systems, already briefly discussed
in the previous section, also neutrinos are being
considered~\cite{ahlucpt,muracpt,nickcpt}
as a possible laboratory for tests of Planck-scale departures
from CPT symmetry.

\subsection{Distance fuzziness}
As one last example of effect that one could plausibly
expect from quantum gravity, I consider here ``distance fuzzyness''.
Once again one is exploring the possibility that some ideas
from quantum theory would apply to spacetime physics.
A key characteristic of quantum theory is the emergence
of uncertainties, and one might expect that the ``distance observable''
would also be affected by uncertainties.
Actually various heuristic arguments suggest that for such
a  ``distance observable'' the uncertainties might be more
pervasive: in ordinary quantum theory one is still able to
measure sharply any given observable, though at the cost
of renouncing all information on a conjugate observable,
but it appears plausible that a  quantum-gravity ``distance observable''
would be affected by irreducible uncertainties.
Most authors would consider a $\delta D \ge L_p$ relation,
meaning that the uncertainty in the measurement of distances
could not be reduced below the Planck-length level,
but measurability bounds of other forms, generically of
the type  $\delta D \ge f(D,L_p)$ (with $f$ some function
such that $f(D,0)=0$) are also being considered.

The presence of such an irreducible measurement uncertainty
could be significant in various contexts.
For example, these ideas would suggest that
the noise levels in the readout of a laser interferometer
would receive an irreducible (fundamental) contribution
from quantum-gravity effects.
Interferometric noise can in principle be reduced to zero
in classical physics, but already the inclusion in the analysis
of the ordinary quantum properties of matter introduces
an extra noise contribution with respect to classical physics.
A fundamental Planck-scale-induced
uncertainty in the length of the arms of the interferometer
would introduce another source of noise,
and the possibility of testing this idea is presently
under investigation
(see, {\it e.g.}, Refs.~\cite{gacgwi,bignapap,nggwi,bignapatwo}).

\subsection{Aside on the differences between systematic
and nonsystematic effects}
It is perhaps useful
to stress the differences between systematic
and nonsystematic Planck-scale effects,
which I can illustrate using the
the type of effects discussed in the previous parts
of this section.

An example of systematic effect is given by the departures
from Lorentz symmetry encountered in certain noncommutative
spacetimes (on which I shall return later in these lectures).
There the Planck-scale structure of spacetime can introduce
a systematic dependence of the speed of photons on their wavelength.
After a journey of duration $T$ the difference between the expected
position of the photon and the Planck-scale-corrected position
could take the form $\Delta x \sim T \delta v \sim c T L_p/\lambda$,
where $\lambda$ is the photon wavelength.

If we instead focus on how ``distance fuzziness'' could affect
the propagation of photons it is natural to expect that
a group of photons would all travel the same average distance
in a given time $T$
(and this average distance is still given by $c T$),
but for each individual photon the distance travelled
might be slightly different from the average,
as a result of distance fuzziness.
This is an example of nonsystematic effect.
Just to be more specific let us imagine that distance fuzziness effectively
introduces a Planck-length uncertainty in position per each Planck time
of travel. Then the final position uncertainty
would be of the type $\Delta x \sim \sqrt{cT L_p}$.
The square root here (assuming a random-walk-type description)
is the result of the fact that nonsystematic effects do not add linearly,
but rather according to rules familiar in the analysis of stochastic
processes.

\section{A prototype exercise: modified dispersion relations}
In the previous sections I tried to give a general, but rough,
description of how one works in quantum-gravity phenomenology.
I will now discuss a specific example of quantum-gravity-phenomenology
study, with the objective of illustrating in more detail
the type of challenges that one must face and some strategies that
can be used.
The example I am focusing on is the one of Planck-scale modifications
of the energy-momentum dispersion relation, which has been
extensively studied from the quantum-gravity-phenomenology perspective.
I will start with a brief description of how modified dispersion
relations arise\footnote{I discuss
noncommutative spacetimes
and the Loop Quantum Gravity approach, which are the
best understood Planck-scale frameworks in which
it appears that the dispersion relation is Planck-scale modified.
But other types of intuitions about the quantum-gravity problem
may lead to modified dispersion relations, including some realizations
of the idea of ``spacetime foam''~\cite{grbgac,garaytest,emn},
which allow an analogy with the laws of particle
propagation in a thermal environment~\cite{grbgac,garaytest,gacpi}.}
in the study of noncommutative spacetimes
and in the study of loop quantum gravity.
I will then discuss some test theories which might play a
special role in the development of the relevant phenomenology.
And finally I will discuss some observations in astrophysics which
can be used to set limits on the test theories.

\subsection{Modified dispersion relations in
canonical noncommutative spacetime}
The noncommutative spacetimes in which modifications of the dispersion
relation are being most actively considered all fall within
the following rather general parametrization of
noncommutativity of the spacetime coordinates:
\begin{equation}
\left[x_\mu,x_\nu\right] = i \theta_{\mu \nu}
+ i \rho^\beta_{\mu \nu} x_\beta ~.
\label{all}
\end{equation}
It is convenient to first focus on the special case $\rho = 0$,
the ``canonical noncommutative spacetimes''
\begin{equation}
\left[x_\mu,x_\nu\right] = i \theta_{\mu \nu}
~.
\label{cano}
\end{equation}

Of course, the natural first guess for introducing dynamics
in these spacetimes is a quantum-field-theory formalism.
And indeed, for the special case $\rho = 0$,
an approach to the
construction of a quantum field theory has been developed rather
extensively\cite{susskind,dougnekr}.
While most aspects of these field theories closely resemble their
commutative-spacetime counterparts, a surprising feature
that emerges is the
so-called ``IR/UV mixing"\cite{susskind,dougnekr,gianlucaken}:
the high-energy sector of the theory does not decouple
from the low-energy sector.
Connected with this IR/UV mixing is the type of modified dispersion
relations that one encounters in field theory on canonical noncommutative
spacetime, which in general take the form
\begin{equation}
 m^2 \simeq E^2 - \vec{p}^2
+ \frac{\alpha_1}{p^\mu \theta_{\mu \nu} \theta^{\nu \sigma} p_\sigma}
+ \alpha_2
m^2 \ln \left( p^\mu \theta_{\mu \nu} \theta^{\nu \sigma} p_\sigma \right)
+ \dots
~,
\label{dispCANO}
\end{equation}
where the $\alpha_i$ are parameters, possibly taking different
values for different particles (the dispersion relation is not ``universal''),
that depend on various aspects
of the field theory, including its field content and the nature
of its interactions.
The fact that this dispersion relation can be singular in the infrared
is a result of the IR/UV mixing. A part of the infrared singularity
could be removed by introducing (exact) supersymmetry, which typically
leads to $\alpha_1 = 0$.

The implications
of this IR/UV mixing for dynamics are still not fully understood,
and there is still justifiable skepticism\cite{skepticNCFT}
toward the correctness of
the type of field-theory construction adopted so far.
I think it is legitimate to even wonder whether
a field-theoretic formulation of the dynamics is at all truly compatible
with the canonical spacetime noncommutativity.
The Wilson decoupling between IR and UV degrees of freedom
is a crucial ingredient of most applications of field
theory in physics, and it is probably incompatible with
canonical noncommutativity: the associated uncertainty
principle of the type $\Delta x_\mu \Delta x_\nu \ge \theta_{\mu \nu}$
implies that it is not possible to probe short distances
(small, say, $\Delta x_1$) without probing simultaneously
the large-distance regime ($\Delta x_2 \ge \theta_{2,1 }/\Delta x_1$).

In any case, the presence of modified dispersion relations
in canonical noncommutative spacetime should be expected, since
Lorentz symmetry is ``broken'' by the tensor $\theta_{\mu \nu}$.
An intuitive characterization of this Lorentz-symmetry breaking
can be obtained by looking at wave exponentials. The Fourier
theory in canonical noncommutative spacetime is based~\cite{wessLANGUAGE}
on simple wave exponentials $e^{i p^\mu x_\mu}$ and from
the $[x_\mu,x_\nu] = i \theta_{\mu \nu}$
noncommutativity relations one finds that
\begin{equation}
e^{i p^\mu x_\mu} e^{i k^\nu x_\nu}
= e^{-\frac{i}{2} p^\mu
\theta_{\mu \nu} k^\nu} e^{i (p+k)^\mu x_\mu} ~,
\label{expprodcano}
\end{equation}
{\it i.e.} the Fourier parameters $p_\mu$ and $k_\mu$ combine just as
usual, but there is the new ingredient of the overall $\theta$-dependent
phase factor.
The fact that momenta combine in the usual way reflects the fact that
the transformation rules for energy-momentum from one
(inertial) observer to another are still the familiar, undeformed,
Lorentz transformation rules. However, the product of wave exponentials
depends on $p^\mu \theta_{\mu \nu} k^\nu$; it depends on the ``orientation"
of the energy-momentum vectors $p^\mu$ and $k^\nu$
with respect to the $\theta_{\mu \nu}$ tensor.
The $\theta_{\mu \nu}$ tensor plays the role of a
background that
identifies a preferred class of inertial observers\footnote{Note that
these remarks apply to canonical noncommutative spacetimes
as studied in the most recent (often String-Theory inspired) literature,
in which $\theta_{\mu \nu}$ is indeed simply a tensor (for a given
observer, an antisymmetric matrix of numbers).
I should stress however that the earliest studies of canonical noncommutative
spacetimes (see Ref.~\cite{dopl1994} and follow-up work)
considered a $\theta_{\mu \nu}$ with richer mathematical properties,
notably with nontrivial algebra relations with the spacetime coordinates.
In that earlier, and more ambitious, setup it is not obvious that Lorentz
symmetry would be broken: the fate of Lorentz symmetry
may depend on the properties (dynamics?)
attributed to $\theta_{\mu \nu}$.}.
Different particles can be affected by the presence of this background
in different ways, leading to the emergence of different
dispersion relations.
All this is consistent with indications of the mentioned
popular field theories in canonical noncommutative spacetimes.

\subsection{Modified dispersion relations in
 $\kappa$-Minkowski noncommutative spacetime}
In canonical noncommutative spacetimes Lorentz symmetry is ``broken''
and there is growing evidence that Lorentz symmetry breaking occurs
for most choices of the tensors $\theta$ and $\rho$.
It is at this point clear, in light of several recent results,
that the only way to preserve Lorentz symmetry
is the choice $\theta = 0 =\rho $, {\it i.e.} the case in which
there is no noncommutativity
and one is back to the familiar classical
commutative Minkowski spacetime.
When noncommutativity is present
Lorentz symmetry is usually
broken, but
recent results suggest that for some special choices of the
tensors $\theta$ and $\rho$
Lorentz symmetry might be deformed,
in the sense of the recently \index{Doubly Special Relativity}
proposed ``doubly-special relativity''
scenario~\cite{gacdsr}, rather than broken.
In particular, this appears to be the case for the Lie-algebra
$\kappa$-Minkowski~\cite{majrue,kpoinap,gacmajid,lukieFT,gacmich,wesskappa}
noncommutative spacetime ($l,m = 1,2,3$)
\begin{equation}
\left[x_m,t\right] = {i \over \kappa} x_m ~,~~~~\left[x_m, x_l\right] = 0 ~.
\label{kmindef}
\end{equation}

$\kappa$-Minkowski
is a Lie-algebra spacetime that clearly enjoys classical space-rotation
symmetry; moreover, at least in a Hopf-algebra sense (see, {\it e.g.},
Ref.~\cite{gacmich}), $\kappa$-Minkowski
is invariant under ``noncommutative translations''.
Since I am focusing here on Lorentz symmetry,
it is particularly noteworthy that in $\kappa$-Minkowski
boost transformations are necessarily modified~\cite{gacmich}.
A first hint of this comes from the necessity of a deformed
law of composition of momenta, encoded
in the so-called coproduct  \index{co-product}
(a standard structure for a Hopf algebra).
One can see this clearly by considering the Fourier tranform.
It turns out~\cite{gacmajid,lukieFT} that in the
$\kappa$-Minkowski case the correct formulation of the Fourier theory
requires a suitable ordering prescription
for wave exponentials. From
\begin{equation}
 :e^{i k^\mu x_\mu}: \equiv e^{i k^m x_m} e^{i k^0 x_0}
~,
\label{order}
\end{equation}
as a result of $[x_m,t] = i x_m/\kappa$
(and $[x_m, x_l] = 0$),
it follows that
the wave exponentials combine in a nontrivial way:
\begin{equation}
(:e^{i p^\mu x_\mu}:) (:e^{i k^\nu x_\nu}:) =
:e^{i (p \dot{+} k)^\mu x_\mu}:
\quad.
\label{expprodlie}
\end{equation}
The notation ``$\dot{+}$" here introduced reflects the
behaviour of the mentioned ``coproduct"
composition of momenta:
\begin{equation}
p_\mu \dot{+} k_\mu \equiv \delta_{\mu,0}(p_0+k_0) + (1-\delta_{\mu,0})
(p_\mu +e^{\lambda p_0} k_\mu) ~. \label{coprod}
\end{equation}

As argued in Refs.~\cite{gacdsr} the nonlinearity of the law of composition
of momenta might require an absolute (observer-independent) momentum scale,
just like upon introducing a nonlinear law of composition of velocities
one must introduce the absolute observer-independent scale of
velocity $c$. The inverse of the noncommutativity scale $\lambda$
should play the role of this absolute momentum scale.
This invites one to consider the possibility~\cite{gacdsr}
that the transformation laws for energy-momentum
between different observers would have two invariants, $c$ and $\lambda$,
as required \index{Doubly Special Relativity}
in ``doubly-special relativity''~\cite{gacdsr}.

On the basis of (\ref{coprod}) one is led~\cite{majrue,kpoinap,gacmajid}
to the following result for the form of the energy/momentum
dispersion relation
\begin{equation}
\left(\frac{2}{\lambda}\sinh\frac{\lambda m}{2}\right)^2 =
\left(\frac{2}{\lambda}\sinh\frac{\lambda E}{2}\right)^2-
e^{\lambda E}\vec{p}^2
~,
\label{dispkpoin}
\end{equation}
which for low momenta takes the approximate form
\begin{equation}
m^2 \simeq E^2 - \vec{p}^2
- \lambda E \vec{p}^2
~.
\label{dispkpoinlimit}
\end{equation}
Actually, the precise form of the dispersion relation
may depend on the choice of ordering prescription
for wave exponentials~\cite{gacmich}
((\ref{dispkpoin}) follows form (\ref{order})),
and this point deserves further scrutiny,
but even setting aside this annoying ordering ambiguity,
there appear to be severe obstructions~\cite{lukieFT,gacmich} for
a satisfactory formulation of a
quantum field theory in $\kappa$-Minkowski.
The techniques that were rather straightforwardly applied
for the construction of field theory in canonical noncommutative spacetime
do not appear to be applicable in the $\kappa$-Minkowski case.
It is not unplausible that the ``virulent'' $\kappa$-Minkowski
noncommutativity may require
some departures from a standard field-theoretic setup.

\subsection{Modified dispersion relation in Loop Quantum Gravity}
Loop Quantum Gravity is one of the most ambitious
approaches to the quantum-gravity
problem, and its
understanding is still in a relatively early stage.
As presently understood, Loop Quantum Gravity predicts an inherently
discretized spacetime~\cite{crLIVING,ashtNEW,leeLQGrev},
and this occurs in a rather compelling way: it is not that one introduces
by hand an {\it a priori} discrete background spacetime; it is rather
a case in which a background-independent analysis ultimately
leads, by a sort of self-consistency, to the emergence of
discretization.
There has been much discussion recently, prompted by the
studies~\cite{grbgac,gampul,mexweave},
of the possibility that this discretization
might lead to broken Lorentz symmetry and a modified dispersion relation.
Although there are
cases in which a discretization is compatible
with the presence of continuous classical
symmetries~\cite{simonecarlo,areanew},
it is of course natural, when adopting a discretized spacetime,
to put Lorentz symmetry under careful scrutiny.
Arguments presented in Refs.~\cite{gampul,mexweave}
suggest that Lorentz
symmetry might indeed be broken in Loop Quantum Gravity.

Moreover, very recently Smolin, Starodubtsev and I proposed~\cite{kodadsr}
(also see the follow-up study in Ref.~\cite{jurekkodadsr})
a mechanism such that Loop Quantum Gravity
would be described at the most fundamental level as a theory that in the
flat-spacetime limit admits deformed Lorentz symmetry,
in the sense of the ``doubly-special relativity''
scenario~\cite{gacdsr}.
Our argument originates from the role that certain quantum symmetry groups
(``q-deformed algebras'')
have in the Loop-Quantum-Gravity description of spacetime with
a cosmological constant, and observing that in the flat-spacetime limit
(the limit of vanishing cosmological constant)
these quantum groups might not contract to a classical Lie algebra,
but rather contract to a quantum (Hopf) algebra.

All these studies point to the presence of a modified dispersion
relation, although different arguments lead to different
intuition for the form of the dispersion relation.
A definite result might have to wait for the solution of
the well-known ``classical-limit problem" of
Loop Quantum Gravity. We are presently
unable to recover from this full quantum-gravity theory the limiting
case in which the familiar quantum-field-theory description of
particle-physics processes in a classical background spacetime applies.
Some recent studies appear to suggest\cite{gampulDENSI}
that in the same contexts in which departures from Lorentz
symmetry may be revealed one should adopt a density-matrix formalism,
and then the whole picture would collapse to the familiar Lorentz-invariant
quantum-field-theory description in contexts involving both relatively
low energies and relatively low boosts with respect to the center-of-mass
frame (e.g. the particle-physics collisions studied at several
particle accelerators).

\subsection{Some issues relevant for the proposal of test theories}
In these lectures I am attempting to stress in particular the
need for quantum-gravity phenomenology to establish that some
Planck-scale pictures of spacetime are falsifiable
and the need to rely on some reference test theories
in the analysis of the progress of experimental limits as
better data become available.

The results I briefly summarized in the previous three subsections
provide a good indication of the fact that falsifiability is
within reach. Both in the analysis of noncommutative spacetimes
and in the analysis of Loop Quantum Gravity there are
a few open issues which do not at present allow us
to describe in detail a falsifiable prediction, but,
in light of the progress achieved over the last few years,
the nature of these open issue encourages us to think that
we should soon achieve falsifiability.

In the meantime quantum-gravity phenomenology will have to
push the limits on the type of effects that are emerging,
and this effort should be guided by the objective of
falsifiability. The analyses should avoid relying on assumptions
which are likely to prove incorrect for the relevant formalisms.
And when the open issues confront us with some alternative scenarios,
the phenomenology work should attempt to ``cover all possibilities'',
 {\it i.e.} push the limits in all directions that are still
 compatible with the present understanding of the formalism
 (so that when the ambiguity is resolved there will be a
 class of data ready for comparison with theory).

In this situation it will be crucial for the development
of the phenomenological programme to adopt some suitably structured
test theories, which should also be useful for bridging the gap
between the experimental data and the, still incomplete, falsifiability
analysis. These test theories should be our common language in assessing
the progresses made in improving the sensitivity of experiments,
a language that must also be suitable for access from the side of
those working at the development of the quantum-gravity/quantum-spacetime
theories.

As we contemplate the challenge of developing such carefully-balanced
test theories it is important to observe that
the most robust part of the results I summarized in the previous
three subsections
is clearly the emergence of modified dispersion relations.
Therefore if one could set up experiments testing
directly the dispersion relation
the resulting limits would have wide applicability.
In principle one could investigate the form of the
dispersion relation directly by making simultaneous
measurements of energy and space-momentum;
however, it is easy to see that achieving Planck-scale
sensitivity in such a direct test is well beyond our capabilities.

Useful test theories on which to base the relevant phenomenology
must therefore combine
the ingredient of the dispersion relation with other ingredients.
As I shall discuss in greater detail later in this section,
there are three key issues for this test-theory development:


\begin{list}{}{}

\item (i) in presence of the modified dispersion relation should
we still assume the validity of
of the relation $v = dE/dp$ between the speed of a particle
and its dispersion relation? (here $dE/dp$ is the derivative of
the function $E(p)$ which of course is implicitly introduced
through the dispersion relation)

\medskip

\item (ii) in presence of the modified dispersion relation should
we still assume the validity of the standard law of
energy-momentum conservation?

\medskip

\item (iii) in presence of the modified dispersion relation
which formalism should be adopted for the description
of dynamics?

\end{list}


The fact that these are key issues is also a consequence
of the type of data that we expect to have access to,
as I shall discuss later in this section.

Unfortunately on these three key points the quantum-spacetime
pictures which are providing motivation for the study
of Planck-scale modifications of the dispersion relation,
reviewed in the previous three subsections,
are not providing much guidance yet.

For example, in Loop Quantum Gravity,
while we do have evidence that the dispersion relation should
be modified, we do not yet have a clear indication concerning
whether the law of energy-momentum conservation should also
be modified and we also cannot yet robustly establish whether
the relation $v=dE/dp$ should be preserved.
Moreover, perhaps most importantly,
some recent studies~\cite{gampulDENSI}
invite us to consider the possibility
that in the same contexts in which
Loop-Quantum-Gravity departures from Lorentz
symmetry may be revealed one should also adopt a density-matrix formalism,
and then the whole picture might reduce to the familiar Lorentz-invariant
quantum-field-theory description in contexts involving both relatively
low energies and relatively low boosts with respect to the center-of-mass
frame. We should therefore be prepared for surprises in the description
of dynamics.

Similarly in the analysis of noncommutative
spacetimes we are close to establishing in
rather general terms that some modification of the dispersion relation
is inevitable,
but other aspects of the framework have not yet been clarified.
While most of the literature for canonical noncommutative spacetimes
assumes~\cite{susskind,dougnekr}
that the law of energy-momentum conservation should not be modified,
most of the literature for $\kappa$-Minkowski spacetime
argues in favour of a modification
(perhaps consistent with the corresponding
doubly-special-relativity criteria~\cite{gacdsr})
of the law of energy-momentum conservation.
There is also still no consensus
on the relation between speed and dispersion relation,
and particularly in the $\kappa$-Minkowski literature
some departures from the $v=dE/dp$ relation are actively
considered~\cite{Kosinski:2002gu,Mignemi:2003ab,Daszkiewicz:2003yr,jurekREV}.
And concerning the formalism to be used for the description
of dynamics in a noncommutative spacetime, while everybody's first
guess is the field-theoretic formalism, the fact that
attempts at a field theory formulation encounter so many difficulties
(the IR/UV mixing for the canonical-noncommutative spacetime case
and the even more pervasive shortcomings of the proposals for a
field theory in $\kappa$-Minkowski)
must invite one to consider possible alternative formulations
of dynamics.

Clearly the situation on the theory side invites us to be prudent:
if a given phenomenological picture relies on too many assumptions
on Planck-scale physics it is likely that it might not reproduce
any of the mentioned quantum-gravity and/or quantum-spacetime models
(when these models are eventually fully understood they will
give us their own mix of Planck-scale features, which is difficult to
guess at the present time).
On the other hand it is necessary for the robust development
of a phenomenology to adopt well-defined test theories.
Without reference to a well-balanced set of test theories
it is impossible to compare the limits obtained in different
experimental contexts, since each experimental context may require
different ``ingredients" of Planck-scale physics.
And it is of course meaningless to compare limits obtained
on the basis of
different conjectures for the Planck-scale regime, especially since
our very limited understanding of the Planck scale regime
should encourage us to be prudent when formulating any assumption
(virtually any assumption about the Planck-scale regime could
turn out to be incorrect, once theories are better understood).

\subsection{A test theory for pure kinematics}
The majority (see, {\it e.g.},
Refs.~\cite{billetal,kifu,ita,aus,gactp})
of studies concerning Planck-scale modifications of the
dispersion relation adopt the phenomenological formula
\begin{equation}
 m^2 \simeq E^2 - \vec{p}^2
+  \eta \vec{p}^2 \left({E^n \over E^n_{p}}\right)
+ O({E^{n+3} \over E^{n+1}_{QG}})
~,
\label{displeadbis}
\end{equation}
with real $\eta$ of order $1$ and integer $n$.
This formula is compatible with some
of the results obtained in the Loop-Quantum-Gravity approach
and reflects the results obtained in $\kappa$-Minkowski and
other noncommutative spacetimes (but, as mentioned, in the special case
of canonical noncommutative spacetimes one encounters
a different, infrared singular, dispersion relation).

As stressed above, on the basis of the status on the theory side,
a prudent approach in combining the dispersion relation
with other ingredients is to be favoured.
Since basically all experimental situations will involve
some aspects of kinematics that go beyond the dispersion
relation (while there are some cases in which the dynamics,
the interactions among particles, does not play a role),
and taking into account the mentioned difficulties in establishing
what is the correct formalism for the description
of dynamics\footnote{I am here using the expression ``dynamics at
the Planck scale'' with some license.
Of course, in our phenomenology we will not be sensitive directly
to the dynamics at the Planck scale. However, as I discuss
in greater detail in the next subsection, if the arguments
that encourage the use of new descriptions of dynamics at
the Planck scale are correct, then a sort of ``order of limits problem''
clearly arises. Our experiments will involve energies much lower
than the Planck scale, and we know that in the infrared limit
the familiar formalism with field-theoretic description of
dynamics and Lorentz invariance will hold. So we would need to
establish whether experiments that are sensitive to Planck-scale
departures from Lorentz symmetry could also be sensitive to
Planck-scale departures from the field-theoretic description
of dynamics. Since we still know very little about this
alternative descriptions of dynamics a prudent approach,
avoiding any assumption about the description of dynamics
is certainly preferable.}
at the Planck scale,
most authors prefer to prudently combine the dispersion relation
with other ``purely kinematical" aspects of Planck-scale physics.

Already in the first studies\cite{grbgac,aemn1}
that proposed a phenomenology
based on (\ref{displeadbis}) it was assumed
that even at the Planck scale the familiar
description of ``group velocity", obtained from the dispersion relation
according to $v=dE/dp$, should hold\footnote{As mentioned,
this assumption is not guaranteed to apply to the formalisms
of interest, and indeed several authors
have considered
alternatives~\cite{Kosinski:2002gu,Mignemi:2003ab,Daszkiewicz:2003yr,jurekREV}.
While the studies advocating
alternatives to $v = dE/dp$ rely of a large variety of
arguments (some more justifiable some less),
in my own perception~\cite{gianluFranc}
a key issue here is whether quantum gravity leads to a modified
Heisenberg uncertainty principle, $ [x,p] =1 + F(p)$. Assuming a Hamiltonian
description is still available, $v = dx/dt \sim [x,H(p)]$,
the relation $v = dE/dp$ essentially follows from $ [x,p] =1$.
But if $ [x,p] \neq 1$ then  $v = dx/dt \sim [x,H(p)]$ would not lead
to $v = dE/dp$. And there is much discussion in the quantum-gravity
community of the possibility of
modifications of the Heisenberg uncertainty principle at the Planck scale.}.

In other works motivated by the analysis reported in Ref.~\cite{grbgac}
another key kinematical feature was introduced:
starting with the studies reported in Refs.~\cite{kifu,ita,aus,gactp}
the dispersion relation (\ref{displeadbis}) and
the velocity relation $v = dE/dp$
were combined with the assumption that the law of energy-momentum conservation
should not be modified at the Planck scale, so that, for example,
in a $a + b \rightarrow c + d$ particle-physics process one would have
\begin{equation}
E_a + E_b = E_c + E_d
~,
\label{econs}
\end{equation}
\begin{equation}
\vec{p}_a + \vec{p}_b = \vec{p}_c + \vec{p}_d
~.
\label{pcons}
\end{equation}


Most authors work within this kinematic framework assuming ``universality''
of the dispersion relation (on which I shall return in the next subsection),
but some have allowed~\cite{jaco,nycksync}
for a particle-dependence and possibly an
helicity-polarization dependence of the coefficients $\eta$, $n$ of
the dispersion relation.

The elements I described in this subsection compose a
quantum-gravity phenomenology test theory
that has already been widely considered in
the literature, even though it was never previously
characterized in detail.
In the following I will refer to this test theory as the ``AEMNS test
theory"\footnote{I am using ``AEMNS'' on the basis of the initials of the
names of the authors in Ref.\cite{grbgac}, which first proposed a
phenomenology based on the dispersion relation (\ref{displeadbis}).
But as mentioned the full test theory, as presently used in most
studies, only emerged gradually in follow-up work. In particular,
there was no discussion of energy-momentum conservation
in Ref.~\cite{grbgac}. Unmodified energy-momentum conservation
was introduced in Refs.~\cite{kifu,ita,aus,gactp}.},
and I will assume that experimental
bounds on this test theory should be placed by using only
the following assumptions:

(AEMNS.1) the dispersion relation is of the form
\begin{equation}
 m^2 \simeq E^2 - \vec{p}^2
+  \eta_a \vec{p}^2 \left({E^{n_a} \over E^{n_a}_{p}}\right)
+ O({E^{n+3} \over E^{n+1}_{QG}})
~,
\label{displeadNONUNI}
\end{equation}
where $\eta_a$ and $n_a$ can in general take different values for
different particles and for different
helicities-polarizations of the same particle
(the index spans over particles and helicities/polarizations);

(AEMNS.2) the velocity of a particle can be obtained from
the dispersion relation using $v=dE/dp$;

(AEMNS.3) the law of energy-momentum conservation is not modified;

(AEMNS.4) nothing is assumed about dynamics ({\it i.e.}
the analysis of this test theory will be limited to classes of
experimental data that involve pure kinematics, without any
role for dynamics).


\subsection{The minimal AEMNS test theory}
On the basis of the results we presently have, at least within
Loop Quantum Gravity and the study of certain noncommutative spacetimes,
the formulation of the ``AEMNS test theory'' discussed in
the previous subsection is general enough
that we should expect it to be relevant for most quantum-spacetime pictures
in which Lorentz symmetry is broken.
Since, as mentioned, the analysis of these models is still in progress
we might eventually be forced to consider further generalizations,
including a possible modifications of the energy-momentum
conservation law\footnote{In a doubly-special-relativity framework
with modified dispersion relation the law of energy-momentum conservation
must be {\underline{correspondingly modified}} in order to preserve
the equivalence of inertial observers~\cite{gacdsr}.
Instead in a framework in which Lorentz symmetry is actually broken,
with the associated loss of equivalence among inertial observers,
modifications of the dispersion relation are in principle compatible
with an unmodified law of energy-momentum conservation. Still, even
in the broken-Lorentz-symmetry case, a modification of the law
of energy-momentum conservation is possible.}
and/or of the law $v=dE/dp$.

Rather then prematurely considering this possible even wider parameter
space, at present it is
more reasonable to focus on a ``minimal version''
of the AEMNS test theory, in which the universality of the dispersion
relation is assumed. It is in fact natural to expect that universality
will be preserved in most of the relevant quantum-spacetime pictures.
Moreover,
as long as this minimal AEMNS test theory
is not ruled out, clearly its more general nonuniversal version
discussed in the previous subsection cannot be ruled.
And it will be very useful to have a simple two-parameter space
to use as reference in keeping track of the gradual improvement
of the experimental sensitivities.

In order to be self-contained
let me list here the characteristics
of this ``minimal AEMNS test theory'':

(minAEMNS.1) the dispersion relation is of the form
\begin{equation}
 m^2 \simeq E^2 - \vec{p}^2
+  \eta \vec{p}^2 \left({E^{n} \over E^{n}_{p}}\right)
+ O({E^{n+3} \over E^{n+1}_{QG}})
~,
\label{displeadNONUNImin}
\end{equation}
where $\eta$ and $n$ are universal (same value for every particle
and for both helicities/polarizations of a given particle);

(minAEMNS.2) the velocity of a particle can be obtained from
the dispersion relation using $v=dE/dp$;

(minAEMNS.3) the law of energy-momentum conservation is not modified;

(minAEMNS.4) nothing is assumed about dynamics ({\it i.e.}
the analysis of this test theory will be limited to classes of
experimental data that involve pure kinematics, without any
role for dynamics).

\subsection{A test theory based on low-energy effective field theory}
The AEMNS test theory has the merit of relying only on a relatively
small network of assumptions on kinematics, without assuming
anything about the role of the Planck scale in dynamics.
However, of course, this justifiable prudence turns into
a severe limitation on the class of experimental contexts
which can be used to constrain the parameters of the test theory.
It is in fact rather rare that a phenomenological analysis
can be completed without using (more or less explicitly)
any aspects of the interactions among the particles involved
in the relevant processes.
The desire to be able to analyze a wider class of experimental contexts
is therefore providing motivation for the development of
test theories more ambitious than the AEMNS test theory,
with at least some elements of dynamics.
This is understandable but, in light of the situation on the theory
side, work with one of these more ambitious test theories
should proceed with the awareness that there is a high risk that
it may turn out that
none of the quantum-gravity approaches which are being pursued
is reflected in the test theory.

One reasonable possibility to consider, when the urge to analyze data
that involve some contamination from dynamics cannot be resisted,
is the one of describing dynamics within the framework
of low-energy effective field theory.
In this subsection I want to discuss a test theory
which is indeed based on low-energy effective field theory,
and has emerged from the work recently reported in
Ref.~\cite{rob} (which is rooted in part in the earlier Ref.~\cite{gampul}).

Before a full characterization of this test theory
I should first warn the reader that there might be some
severe limitations for the applicability of
low-energy effective field theory to the investigation of
Planck-scale physics,
especially when departures from Lorentz symmetry are present.

A significant portion of the quantum-gravity community
is in general, justifiably, skeptical about the results obtained
using low-energy effective field theory in
analyses relevant for the quantum-gravity problem.
After all the first natural prediction of
low-energy effective field theory
in the gravitational realm is a value of the energy density
which is some 120 orders of magnitude greater
than allowed by observations\footnote{And the
outlook of low-energy effective field theory
in the gravitational realm does not improve
much through the observation
that exact supersymmetry could protect from
the emergence of any energy density.
In fact, Nature clearly does not have
supersymmetry at least
up to the TeV scale,
and this would still lead to a natural prediction
of the cosmological constant
which is some 60 orders of magnitude too high.}.
Somewhat related to this ``cosmological constant problem"
is the fact that a description of possible Planck-scale departures from
Lorentz symmetry within effective field theory can only be developed
with a rather strongly pragmatic attitude; in fact,
while one can introduce Planck-scale suppressed effects
at tree level, one
expects that loop corrections would typically lead to inadmissibly large
departures from ordinary Lorentz symmetry.
Indeed some studies, notably Refs.~\cite{suda1,suda2},
have shown mechanisms such that, within an effective-field-theory
formulation, loop effects would lead to
inadmissibly large
departures from ordinary Lorentz symmetry,
which could be avoided only by introducing a large level
of fine tuning.

It is rather amusing that alongside with numerous
researchers who are skeptical about any results obtained
using low-energy effective field theory in
analyses relevant for the quantum-gravity problem,
there are also quite a few researchers
interested in the quantum-gravity problem
who are completely serene in assuming
that all quantum-gravity effects
should be describable in terms of effective field theory
in low-energy situations.
The (quasi-)rationale behind this assumption is that
field theory works well at low energies without gravity, and quantum gravity
of course must reproduce field theory in an appropriate limit,
so one might expect that at least at low energies the quantum-gravity
effects could be described in the language of field theory as
correction terms to be added to standard lagrangians.

I feel that, while of course an effective-field-theory
description may well turn out to be correct in the end,
the {\it a priori} assumption
that a description in terms of
effective low-energy field-theory should work is rather naive.
If the arguments
that encourage the use of new descriptions of dynamics at
the Planck scale are correct, then a sort of ``order of limits problem''
clearly arises. Our experiments will involve energies much lower
than the Planck scale, and we know that in some limit (a limit
that characterizes our most familiar observations)
the field-theoretic description
and Lorentz invariance will hold. So we would need to
establish whether experiments that are sensitive to Planck-scale
departures from Lorentz symmetry could also be sensitive to
Planck-scale departures from the field-theoretic description
of dynamics.
As an example, let me mention the possibility (not unlikely in
a context which is questioning the fate of Lorentz symmetry)
that quantum gravity would admit a field-theory-type description
only in reference frames in which the process of interest
is essentially occurring in its center of mass
(no ``Planck-large boost''~\cite{mg10qg1}
with respect to center-of-mass frame).
The field theoretic description could emerge in
a sort of ``low-boost limit'', rather than the expected
low-energy limit.
The regime of low boosts with respect the center-of-mass frame
is often indistinguishable with respect to the low-energy limit.
For example, from a Planck-scale perspective, our laboratory
experiments (even the ones conducted at, {\it e.g.} CERN, DESY, SLAC...)
are both low-boost (with respect to the center of mass frame)
and low-energy.
However, the ``UHE cosmic-ray paradox'', for which
a quantum-gravity origin has been conjectured (see later),
occurs in a situation where all the energies of the particles
are still tiny with respect to the Planck energy scale,
but the boost with respect to the center-of-mass frame
(as measured by the ratio $E/m_{proton}$ between the proton energy and
the proton mass) could be considered to be ``large''
from a Planck-scale perspective ($E/m_{proton} \gg E_p/E$).

These concerns are strengthen by looking at the literature
available on  the quantum pictures of spacetime that provide
motivation for the study of modified dispersion relations,
which usually involve
either noncommutative geometry or Loop Quantum Gravity,
where, as mentioned, the outlook of a
low-energy effective-field-theory description is not encouraging.
The construction of field theories
in noncommutative spacetimes requires
the introduction of several new technical tools, which in turn
lead to the emergence of several new physical
features, even at low energies.
I guess that these difficulties arise from the fact
that a spacetime characterized by an uncertainty relation
of the type $\delta x \, \delta y \ge \theta (x,y)$
never really behaves has a classical spacetime, not even at very low energies.
In fact, some low-energy processes will involve soft momentum exchange
in the $x$ direction (large $\delta x$) which however is connected
with the exchange of a hard momentum in the $y$ direction
($\delta y \ge \theta/\delta x$), and this feature cannot be faithfully
captured by our ordinary field-theory formalisms.
In the case of canonical noncommutative spacetimes
one does obtain a plausible-looking field theory\cite{dougnekr},
but the results actually show that it is not possible to rely
on an ordinary effective low-energy
quantum-field-theory description. In fact,
the ``IR/UV mixing"\cite{susskind,dougnekr,gianlucaken}
is such that the high-energy sector of the theory does not decouple
from the low-energy sector, and this in turn affects
very severely\cite{gianlucaken}
the outlook of analyses based on
an ordinary effective low-energy quantum-field-theory description.
For other (non-canonical) noncommutative spacetimes
we are still struggling in the search of a satisfactory formulation of
a quantum field theory~\cite{lukieFT,gacmich}, and it is at this point
legitimate to suspect that such a formulation of dynamics in those
spacetimes does not exist.

Incidentally let me observe that the issues encountered in
dealing with the IR/UV mixing may be related
to my concerns about the large-boost limit of quantum gravity.
In a theory with IR/UV mixing nothing peculiar might be expected for,
say, a collision between two photons both of $MeV$ energy,
but the boosted version of this collision, where one photon
has, say, energy of $100 TeV$ and the other photon
has energy of $10^{-2} eV$, could be subject to
the IR/UV mixing effects, and be essentially untreatable
from a low-energy effective-field-theory perspective.

And noncommutative spacetimes are not the only cases where
an ordinary field-theory description may be inadequate.
As mentioned, the assumption of availability of
an ordinary effective low-energy quantum-field-theory description
finds also no support in Loop Quantum Gravity.
Indeed, so far, in Loop Quantum Gravity
all attempts to find a suitable limit of the theory which can be described
in terms of a quantum-field-theory in background spacetime
have failed.
And on the basis of the recent results of Ref.~\cite{gampulDENSI}
it appears plausible that in several contexts in which one
would naively expect a low-energy field theory description
Loop Quantum Gravity might instead require a density-matrix
description.

Of course, in phenomenology this type of concerns can be set aside, since
one is primarily looking for confrontation with experimental data,
rather than theoretical prejudice. It is clearly legitimate to set up
a test theory exploring the possibility of Planck-scale departures
from Lorentz symmetry within the formalism of low-energy effective
field theory. But one must then keep in mind that the implications
for most quantum-gravity research lines of the experimental bounds
obtained in this way might be very limited.
This will indeed be the case if we discover that, as some mentioned
preliminary results suggest,
the limit in which the full quantum-gravity theory reproduces a
description in terms of effective field theory in classical spacetime
is also the limit in which the departures from Lorentz symmetry
must be neglected.

Having provided this long warning, let me now proceed to
a characterization of the test theory which I see emerging from
the works reported in Refs.~\cite{rob,gampul}.
These studies explore the possibility of a linear-in-$L_p$
modification of the dispersion relation
\begin{equation}
 m^2 \simeq E^2 - \vec{p}^2
+  \eta \vec{p}^2 L_p E
~,
\label{dispROB}
\end{equation}
 {\it i.e.} the case $n=1$ of Eq.~(\ref{displeadbis}).
The key assumption in Refs.~\cite{rob,gampul} is that such
modifications of the dispersion relation should be introduced
consistently with an effective low-energy field-theory description
of dynamics.
The implications of this assumption were explored in particular
for fermions and photons.
It became quickly clear that in such a setup universality cannot
be assumed, since one must at least accommodate a polarization
dependence for photons: in the field-theoretic setup it turns
out that when right-circular polarized photons satisfy the
dispersion relation $E^2 \simeq p^2 + \eta_\gamma p^3$ then necessarily
left-circular polarized photons satisfy the ``opposite sign"
dispersion relation $E^2 \simeq p^2 - \eta_\gamma p^3$.
For spin-$1/2$ particles the analysis reported in
Ref.~\cite{rob} does not necessarily suggest a
similar helicity dependence, but of course in a context in
which photons experience such a complete correlation of the
sign of the
effect with polarization
it would be awkward to assume that instead for
electrons the effect is completely helicity independent.
One therefore introduces two independent parameters $\eta_+$ and $\eta_-$
to characterize the
modification of the dispersion relation for electrons.

In the following I will refer to this test theory as the ``GPMP test
theory" (from the initials of the authors of Refs.~\cite{rob,gampul}),
and I will assume that experimental
bounds on this test theory should be placed by using only
the following assumptions:

(GPMP.1) for right-circular polarized photons
are governed by the dispersion relation
\begin{equation}
 m^2 \simeq E^2 - \vec{p}^2
+  \eta_\gamma \vec{p}^2 \left({E \over E_{p}}\right)
~,
\label{displeadNONUNIgpmp}
\end{equation}
while left-circular polarized photons
are governed by the dispersion relation
\begin{equation}
 m^2 \simeq E^2 - \vec{p}^2
-  \eta_\gamma \vec{p}^2 \left({E \over E_{p}}\right)
~;
\label{displeadNONUNIgpmp2}
\end{equation}

(GPMP.2) for fermions the dispersion relation
takes the form
\begin{equation}
 m^2 \simeq E^2 - \vec{p}^2
+  \eta^{a}_R \vec{p}^2 \left({E \over E_{p}}\right)
~,
\label{displeadNONUNIgpmp3}
\end{equation}
in the positive-helicity case, while
for negative-helicity fermions
\begin{equation}
 m^2 \simeq E^2 - \vec{p}^2
+  \eta^{a}_L \vec{p}^2 \left({E \over E_{p}}\right)
~;
\label{displeadNONUNIgpmp4}
\end{equation}
the index $a$ here reflecting a possible particle dependence;

(GPMP.3) dynamics is described in terms of effective low-energy
field theory.

\subsection{The minimal GPMP test theory}
As in the case of the AEMNS test theory, while a large parameter space should
be considered in order to achieve full generality,
it appears wise to first focus the phenomenology on a reduced version
of the test theory, reflecting some natural physical assumptions.
As in the case of the AEMNS test theory, a reduced
two-parameter space would be ideal for the first-level
description of the gradual improvement
of the experimental sensitivities.
As usual, once the reduced version of the test theory is falsified
one can contemplate its possible generalization (if the developments
on the pure-theory side still justify such an effort from the perspective
of falsification of the theories).

In introducing a reduced GPMP test theory
I believe that a key point of naturalness comes from the
observation that the effective-field-theory setup
imposes for photons a modification of the dispersion relation
which has the same magnitude for both polarizations
but opposite sign: it is then natural to give priority to the
hypothesis that for fermions a similar mechamism would apply, {\it i.e.}
the modification of the dispersion relation should have the same
magnitude for both signs of the helicity, but have a correlation
between the sign of the helicity and the sign of the dispersion-relation
modification.
This would correspond to the natural-looking
assumption that the Planck-scale
effects are such that in a beam composed of randomly selected particles
the average speed in the beam is still governed by ordinary special
relativity (the Planck scale effects average out summing over
polarization/helicity).

A further ``natural'' reduction of the parameter space
is achieved by assuming that all fermions
are affected by the same modification
of the dispersion relation.

The reduced GPMP test theory that emerges from this requirements
is perhaps the most natural among the possible two-parameter
reduction of the GPMP test theory.
In the following I refer to this reduced
GPMP test theory as the ``minimal GPMP test theory''\footnote{Whereas
for the AEMNS test theory there is clearly only one obvious
way to set up the reduction to a two-dimensional parameter space,
within the GPMP test theory, with its automatic polarization dependence
of the effects for photons, one could probably envision more than
one way to set up the reduction to a two-dimensional parameter space.
In a certain sense the two-dimensional parameter space on which I
propose to focus for the AEMNS test theory is {\underline{the}}
minimal AEMNS test theory, whereas here I am
proposing {\underline{a}} minimal GPMP test theory.},
characterized by the following ingredients:

(minGPMP.1) right-circular polarized photons
are governed by the dispersion relation
\begin{equation}
 m^2 \simeq E^2 - \vec{p}^2
+  \eta_\gamma \vec{p}^2 \left({E \over E_{p}}\right)
~,
\label{displeadNONUNIgpmpmin1}
\end{equation}
while left-circular polarized photons
are governed by the dispersion relation
\begin{equation}
 m^2 \simeq E^2 - \vec{p}^2
-  \eta_\gamma \vec{p}^2 \left({E \over E_{p}}\right)
~;
\label{displeadNONUNIgpmpmin2}
\end{equation}

(minGPMP.2) for fermions the dispersion relation
takes the form
\begin{equation}
 m^2 \simeq E^2 - \vec{p}^2
+  \eta_f \vec{p}^2 \left({E \over E_{p}}\right)
~,
\label{displeadNONUNIgpmpmin3}
\end{equation}
in the positive-helicity case, while
for negative-helicity fermions
\begin{equation}
 m^2 \simeq E^2 - \vec{p}^2
-  \eta_f \vec{p}^2 \left({E \over E_{p}}\right)
~,
\label{displeadNONUNIgpmpmin4}
\end{equation}
with the same value of $\eta_f$ for all fermions;

(minGPMP.3) dynamics is described in terms of effective low-energy
field theory.

\subsection{Derivation of limits from analysis of gamma-ray bursts}\label{grbs}
Both in the AEMNS test theory and in the GPMP test theory
one expects a wavelength dependence of the speed of photons,
by combining the modified dispersion relation and
the relation $v = dE/dp$. At ``intermediate energies" ($m < E \ll E_p$)
this velocity law will take the form
\begin{equation}
v \simeq 1 - \frac{m^2}{2 E^2} +  \eta \frac{n+1}{2} \frac{E^n}{E_p^n}
~.
\label{velLIVbis}
\end{equation}
Whereas in ordinary special relativity two photons ($m=0$)
emitted simultaneously would always reach simultaneously a far-away detector,
according to (\ref{velLIVbis}) two simultaneously-emitted
photons should reach the detector at different times
if they carry different energy. Moreover, in the case of the GPMP test
theory even photons with the same energy would arrive at different
times if they carry different polarization.
In fact, while the minimal AEMNS test theory assumes universality,
and therefore
a formula of this type would apply to photons of any polarization,
in the GPMP test theory, as mentioned, the sign of the effect
is correlated with polarization.
As a result, while the AEMNS test theory is best tested by
comparing the arrival times of particles of different energies,
the GPMP test theory is best tested by considering the highest-energy
photons available in the data and looking for a sizeable spread
in times of arrivals (which one would then attribute to
the different speeds of the two polarizations).

This time-of-arrival-difference effect
can be significant\cite{grbgac,billetal}
in the analysis of short-duration gamma-ray bursts \index{gamma-ray bursts}
that reach
us from cosmological distances.
For a gamma-ray burst it is not uncommon that the time travelled
before reaching our Earth detectors be of order $T \sim 10^{17} s$.
Microbursts within a burst can have very short duration,
as short as $10^{-3} s$ (or even $10^{-4} s$), and this
means that the photons
that compose such a microburst are all emitted at the same time,
up to an uncertainty of $10^{-3} s$.
Some of the photons in these bursts
have energies that extend at least up to the $GeV$ range.
For two photons with energy difference of order $\Delta E \sim 1 GeV$
a $\eta \Delta E/E_p$ speed difference over a time of travel
of $10^{17} s$
would lead to a difference in times of arrival of
order
\begin{equation}
\Delta t \sim \eta T \Delta \frac{E}{E_p} \sim 10^{-2} s
~,
\label{tdelay}
\end{equation}
which is significant (the time-of-arrival differences would be larger
than the time-of-emission differences within a single microburst).

For the AEMNS test theory the
Planck-scale-induced time-of-arrival difference
could be revealed\cite{grbgac,billetal}
upon comparison of the ``average arrival time''
 of the gamma-ray-burst signal (or better a microburst within the
 burst)
in different energy channels.
The GPMP test theory would be most effectively tested by looking
for a dependence of the time-spread of the bursts that grows
with energy (at low energies the effect is anyway small, so the
polarization dependence is ineffective, whereas at high energies
the effect may be nonnegligible and an overall time-spread of
the burst could result from the dependence of speed on
polarization).

The sensitivities achievable~\cite{glast}
with the next generation of gamma-ray telescopes,
such as GLAST~\cite{glast},
could allow to test very significantly (\ref{velLIVbis})
in the case $n=1$, by possibly pushing the
limit on $\eta$ far below $1$ (whereas the effects found in the
case $n=2$, $|\eta| \sim 1$ are too small for GLAST).
Whether or not these levels of sensitivity
to the Planck-scale effects are actually achieved
may depend on progress in understanding other aspects of
gamma-ray-burst physics.
In fact, the Planck-scale-effect
analysis would be severely affected if there
were poorly understood at-the-source correlations between
energy of the photons and time of emission.
In the recent Ref.~\cite{piranKARP} it was emphasized that
it appears that one can infer such an energy/time-of-emission
correlation from available gamma-ray-burst data.
The studies of Planck-scale effects will be therefore confronted
with a severe challenge of ``background/noise removal''.
At present it is difficult to guess whether this problem
can be handled successfully. We do have a good card to play
in this analysis: the Planck-scale picture predicts that
the times of arrival should depend on energy in a way that
grows in exactly linear way with the distance of the source.
One may therefore hope that, once a large enough sample
of gamma-ray bursts (with known source distances) becomes available,
one might be able
disentangle the Planck-scale propagation effect
from the at-the-source background.

An even higher sensitivity to possible Planck-scale
modifications of the velocity law could be achieved
by exploiting the fact that, according to
current models~\cite{grbNEUTRINOnew},
gamma-ray bursters should also emit a substantial amount of
high-energy neutrinos.
Some neutrino observatories should soon observe neutrinos with energies
between $10^{14}$ and $10^{19}$ $eV$, and one could, for example, compare
the times of arrival of these neutrinos emitted by
gamma-ray bursters to the corresponding times of arrival of
low-energy photons.
One could use this strategy to test rather stringently\footnote{Note
however that in an analysis mixing the properties of different particles
the sensitivity that can be achieved will depend strongly on whether
universality of the modification of the dispersion relation
is assumed.
For example, for the GPMP test theory
a comparison of times of arrival of neutrinos and photons
could only introduce a bound on some combination of the
dispersion-relation-modification parameters for the photon and for the
neutrino sectors.}
the case
of (\ref{velLIVbis}) with $n=1$, an even perhaps gain some access to
the investigation of the case $n=2$.

In order to achieve these sensitivities with neutrino studies
once again
some technical and conceptual challenges should be
overcome. Also this type of analysis requires
an understanding
of gamma-ray bursters good enough to establish whether there are typical
at-the-source time delays. The analysis would loose much of its potential
if one cannot exclude some systematic tendency of
gamma-ray bursters to emit high-energy
neutrinos with, say, a certain delay with
respect to microbursts of photons. But also in this case
one could hope to combine several observations
from gamma-ray bursters at different distances in order
to disentangle the possible at-the-source effect.

\subsection{Derivation of limits from analysis of UHE cosmic rays}\label{uhecr}
With a given dispersion relation and a given rule for energy-momentum
conservation one has a complete ``kinematic scheme" for the analysis
of the kinematical requirements
for particle production in collisions or decay processes.
Both the AEMNS test theory and the GPMP test theory involve
modified dispersion relations and unmodified laws of energy-momentum
conservation (the fact that the law of energy-momentum conservation
is not modified is explicitly among the ingredients of the AEMNS
test theory, while in the GPMP test theory it follows from
the adoption of low-energy effective field theory).

In these lectures I am not discussing in detail the case of
modified dispersion relations introduced within a ``doubly-special
relativity'' scenario~\cite{gacdsr,dsrmost}.
For clarity of the presentation, I thought it would be best to limit to
two the number of test theories I consider.
Test theories for doubly-special
relativity scenarios with modified dispersion relations
are under consideration (see, {\it e.g.}, Ref.~\cite{dsrphen}),
but I will not make room for them here. It is appropriate however
to stress here that the assumption of
modified dispersion relations and unmodified laws of energy-momentum
conservation is inconsistent with the doubly-special
relativity principles, since it inevitably~\cite{gacdsr} gives rise to
a preferred class of inertial observers.
A doubly-special
relativity scenario with modified dispersion relations must necessarily
have a modified law of energy-momentum conservation.

Going back to the AEMNS and GPMP test theories which I am considering,
in this subsection I want to stress that
combining a modified dispersion relation
with unmodified laws of
energy-momentum conservation
one naturally finds a modification
of the threshold requirements for
certain particle-producing processes.
Let us for example consider, from the AEMNS perspective,
the dispersion relation (\ref{displeadbis}),
with $n=1$, in the analysis of a
collision between
a soft photon of energy $\epsilon$
and a high-energy photon of energy $E$ that creates an electron-positron
pair: $\gamma \gamma \rightarrow e^+ e^-$.
For given soft-photon energy $\epsilon$,
the process is allowed only if $E$ is greater than a certain
threshold energy $E_{th}$ which depends on $\epsilon$ and $m_e^2$.
For $n=1$, combining (\ref{displeadbis}) with unmodified
energy-momentum conservation,
this threshold energy
(assuming $\epsilon \ll m_e \ll E_{th} \ll E_p$)
is estimated as
\begin{equation}
E_{th} \epsilon + \eta \frac{E_{th}^3}{8 E_p}= m_e^2
~.
\label{thrTRE}
\end{equation}
The special-relativistic result $E_{th} = m_e^2 /\epsilon$
corresponds of course to the $\eta \rightarrow 0$ limit
of (\ref{thrTRE}).
For $|\eta | \sim 1$ the Planck-scale correction can be
safely neglected as long as $\epsilon > (m_e^4/E_p)^{1/3}$.
But eventually, for sufficiently small values of $\epsilon$ and
correspondingly large values of $E_{th}$, the
Planck-scale correction cannot be ignored\cite{ita,aus,gactp,jaco,gacpion}

And the process $\gamma \gamma \rightarrow e^+ e^-$
is not the only case in which this type of Planck-scale modification
can be important. There has been strong
interest\cite{kifu,ita,aus,gactp,jaco,gacpion,orfeupion,nguhecr}
in ``photopion production", $p \gamma \rightarrow p \pi$,
where again the combination of (\ref{displeadbis}) with unmodified
energy-momentum conservation leads to a modification of the
minimum proton energy required by the process (for fixed photon energy).
In the case in which the photon energy is the one typical of CMBR photons
one finds that the threshold proton energy can be significantly shifted
upward (for negative $\eta$), and this
in turn should affect at an observably large level the
expected ``GZK cutoff" for the observed cosmic-ray spectrum.
Observations reported by the AGASA\cite{agasa} cosmic-ray
observatory provide some encouragement for the idea of
such an upward shift of the GZK cutoff, but the issue
must be further explored.
Forthcoming cosmic-ray observatories, such as Auger\cite{auger},
should be able\cite{kifu,gactp} to fully investigate this possibility.

In this context the comparison of the AEMNS test theory and the GPMP
test theory is rather straightforward. We are in fact considering a
purely kinematical effect: the shift of a threshold requirement.
For the minimal AEMNS test theory there is a clear prediction that
for negative $\eta$ there should be an upward shift
of the GZK threshold.
For the GPMP test theory one would predict an increase of
the GZK threshold if any one (or both) of the two
helicities of the proton has dispersion
relation of ``negative $\eta$'' type.
If both helicities have dispersion relation of negative-$\eta$ type
then the effect looks rather similar to the corresponding effect
in the AEMNS test theory.
For the situation which I proposed as the ``minimal
GPMP test theory'', where for one of the helicities the dispersion relation
is of negative-$\eta$ type and for the other
helicity the dispersion relation
is of positive-$\eta$ type,
one would expect roughly one half of the UHE protons to evade
the GZK cutoff, so the cutoff would still be violated but in
a softer way than in the case of the AEMNS test theory with
negative $\eta$.

It appears likely that, if the Auger data should actually show evidence
of the expected GZK cutoff,
we would
then be in a position to rule out
the case of negative $\eta$ for the minimal
AEMNS test theory,
and to rule out both the positive-$\eta_f$
and negative-$\eta_f$ case for
the minimal GPMP test theory.
In fact, in the minimal AEMNS test theory
violations of the GZK cutoff are predicted for negative $\eta$
(while they are not present in the positive-$\eta$ case),
while in the minimal GPMP test theory violations of
the GZK cutoff (although less numerous than expected in the
minimal AEMNS test theory with negative $\eta$) are always expected,
independently of the sign of $\eta_f$ (depending on the
sign of $\eta_f$ the protons that violate the GZK cutoff would
have a corresponding helicity).

I should stress that these studies of
the cosmic-ray GZK threshold provide an example in which
the fact that we do not really identify some of the particles in the
relevant particle-physics processes,
an analysis which could in principle be involving pure kinematics,
ends up being exposed to the risk of contamination from
some aspects of dynamics.
If the only background radiation available for photopion production
was the CMBR, then
the prediction of an upward shift of the GZK
cosmic-ray cutoff within the AEMNS test theory,
for negative $\eta$, would be completely robust.
But background radiation has many components and one could
contemplate the possibility to combine AEMNS kinematics
with an unspecified description of dynamics such
that interactions of cosmic rays with other components of the
background radiation would lead to a net result that does not change
the numerical value of the GZK threshold.
While this possibility must be contemplated, I also want to stress that,
at least for $n=1$ and negative $\eta$ of order 1,
this ``conspiracy scenario'' is so unbelievable that
it should be dismissed.
In fact, for $n=1$ and negative $\eta$ of order 1
the AEMNS kinematics allows the interaction of cosmic rays
only with photons of  energy higher than the TeV scale
(see Ref.~\cite{gactp}), and
the density of such high-energy background photons is
so low that, even in a prudent phenomenology,
this ``conspiracy scenario'' can indeed be dismissed.

For the
GPMP test theory there is of course no issue of possible conspiracies,
since the field-theoretic setup allows to evaluate cross sections.

\subsection{Derivation of limits from analysis of photon stability}
As in the case of the GZK cutoff for UHE cosmic rays \index{cosmic rays}
there are several examples in which a given process is allowed
in presence of exact Lorentz symmetry but can be
kinematically forbidden in presence of certain
departures from Lorentz symmetry.
The opposite is also possible: some processes that are
kinematically forbidden in presence of exact
Lorentz symmetry become kinematically allowed
in presence of certain
departures from Lorentz symmetry.
The fact that a process is kinematically allowed of course
does not guarantee that it occurs at an observable rate: it
depends on the laws of dynamics and the amplitudes they predict.

Certain observations in astrophysics,
which allow us to establish
that photons of energies up to $\sim 10^{14}eV$
are not unstable,
can be particularly useful~\cite{jaco,gacpion,orfeupion,seth}
in setting limits on some schemes for departures
from Lorentz symmetry.
Let us for example analyze the process $\gamma \rightarrow e^+ e^-$
from the AEMNS perspective,
using the dispersion relation (\ref{displeadbis}), with $n=1$,
and unmodified energy-momentum conservation.
One easily finds a relation between
the energy $E_\gamma$ of the incoming photon, the opening angle $\theta$
between the outgoing electron-positron pair, and the energy $E_+$ of
the outgoing positron (of course the energy of the outgoing electron
is simply given by $E_\gamma - E_+$).
For the region of phase space with $m_e \ll E_\gamma \ll E_p$
this relation takes the form
\begin{eqnarray}
\cos(\theta) &\! \simeq \!& \frac{E_+ (E_\gamma -E_+) + m_e^2
- \eta  E_\gamma E_+ (E_\gamma -E_+)/E_p}{ E_+ (E_\gamma -E_+)} ~,
\label{gammathresh}
\end{eqnarray}
where $m_e$ is the electron mass.

The fact that for $\eta = 0$ Eq.~(\ref{gammathresh}) would
require $cos(\theta) > 1$ reflects the fact that, if Lorentz symmetry
is preserved, the process $\gamma \rightarrow e^+ e^-$ is kinematically
forbidden. For $\eta < 0$ the process is still forbidden, but for
positive $\eta$ high-energy photons can decay
into an electron-positron pair. In fact,
for $E_\gamma \gg (m_e^2 E_p/|\eta |)^{1/3}$
one finds that
there is a region of
phase space where $\cos(\theta) < 1$, {\it i.e.} there is a physical
phase space available for the decay.

The energy scale $(m_e^2 E_p)^{1/3} \sim 10^{13} eV $ is not
too high for testing, since, as mentioned, in
astrophysics we see photons of energies up to $\sim 10^{14}eV$
that are not unstable (they clearly travel safely some large astrophysical
distances).

Within AEMNS kinematics, for $n=1$ and
positive $\eta$ of order 1, it would have been natural
to expect that such photons with $\sim 10^{14}eV$ energy
would not be stable.
Once again, before claiming that $n=1$ and
positive $\eta$ of order 1 is ruled out, one should be concerned
about possible conspiracies.
The fact that the decay of $10^{14}eV$ photons is allowed
by AEMNS kinematics (for $n=1$ and
positive $\eta$ of order 1) of course does not guarantee
that these photons should rapidly decay. It depends on the relevant
probability amplitude, whose evaluation goes beyond the reach
of kinematics.
I am unable to provide an intuition
for how big of a conspiracy would be needed
to render $10^{14}eV$ photons stable compatibly with
AEMNS kinematics with $n=1$ and $\eta = 1$.
My tentative conclusion is that $n=1$ with
positive $\eta$ of order 1 is ruled out ``up to conspiracies'',
but unlike the case of the GZK-threshold analysis I am unprepared
to argue that the needed conspiracy is truly unbelievable.

For the GPMP test theory the photon stability analysis
is weakened because of other reasons. There one does have the support
of the effective-field-theory descrition of dynamics, and within that
framework one can exclude huge suppression by Planck scale effects
of the interaction vertex needed for $\gamma \rightarrow e^+ e^-$
around $\sim 10^{13}eV$, $\sim 10^{14}eV$.
So the limit-setting effort is not weakened by the absence
of an interaction vertex. However, as mentioned, consistency with
the effective-field-theory setup
requires that the two polarizations of the photon acquire
opposite-sign modifications of the dispersion relation.
We observe in astrophysics some photons
of energies up to $\sim 10^{14}eV$
that are stable over large distances,
but as far as we know those photons could be all,
say, right-circular polarized (or all left-circular polarized).
I postpone a detailed analysis to future work, but let me note here
that there is a region of minimal-GPMP parameter space where both
polarizations of a $\sim 10^{14}eV$
photon are unstable (a subset of the region
with $|\eta_f|>|\eta_\gamma|$). That region of the
minimal-GPMP parameter parameter space
is of course excluded by the photon-stability data.

\subsection{Derivation of limits from analysis of synchrotron radiation}
A recent series of
papers\cite{jacoNATv1,newlimit,jaconature,tedreply,carrosync,nycksync,tedsteck}
has focused on the possibility to set limits on Planck-scale modified
dispersion relations focusing on their implications for synchrotron radiation.
By comparing the content of the first estimates\footnote{Ref.~\cite{jacoNATv1}
is at this point obsolete, since the relevant manuscript
has been revised for the published version\cite{jaconature}
and the recent Ref.~\cite{tedsteck} provides an even more
detailed analysis. It is nevertheless useful to consider
this series of manuscripts \cite{jacoNATv1,jaconature,tedsteck}
as an illustration of how much the outlook of a phenomenological
analysis may change in going from the level of
simplistic order-of-magnitude
estimates to the level of careful comparison with
meaningful test theories.} produced in
this research line~\cite{jacoNATv1}
with the understanding that emerged from follow-up
studies~\cite{newlimit,jaconature,tedreply,carrosync,nycksync,tedsteck}
one can gain valuable insight on the risks involved in analyses based on
simplistic order-of-magnitude
estimates, rather than
careful comparison with
meaningful test theories.
In Ref.\cite{jacoNATv1} the starting point is the observation
that in the conventional (Lorentz-invariant) description of synchrotron
radiation one can estimate the characteristic energy $E_c$ of
the radiation through a heuristic analysis~\cite{jackson}
leading to the formula
\begin{equation}
E_c \simeq {1 \over
R {\cdot} \delta {\cdot} [v_\gamma - v_e]}
~,
\label{omegacjack}
\end{equation}
where $v_e$ is the speed of the electron,
$v_\gamma$ is the speed
of the photon, $\delta$ is an angle obtained from the opening angle between
the direction of the electron and the direction of the
emitted photon, and $R$ is the radius of curvature of
the trajectory of the electron.

Assuming that the only Planck-scale modification in this formula
should come from
the velocity law (described using $v=dE/dp$
in terms of the modified dispersion relation),
one finds that in some instances the characteristic energy of
synchrotron
radiation may be significantly modified by the presence of
Planck-scale departures from Lorentz symmetry.
As an opportunity to test such a modification of the
value of the synchrotron-radiation characteristic energy one
can hope to use some relevant data\cite{jacoNATv1,jaconature}
on photons detected from the Crab nebula. \index{Crab Nebula}
This must be done with caution since
the observational information on synchrotron radiation being emitted
by the Crab nebula is rather indirect: some of the photons we observe
from the Crab nebula are attributed to sychrotron processes on the basis
of a promising conjecture, and the value of the
relevant magnetic fields is also conjectured (not directly measured).

Assuming that indeed the observational situation has been properly
interpreted, and relying on the mentioned assumption that
the only modification to be taken into account is the
one of the velocity law,
one could basically rule out~\cite{jacoNATv1} the case $n=1$
with negative $\eta$
for a modified dispersion relation
of the type (\ref{displeadbis}).

This observation led at first to some excitement,
but more recent papers are starting to adopt a more prudent viewpoint.
The lack of
comparison with a meaningful test theory represents a
severe limitation of the original analysis.
In particular, synchrotron radiation is due to the acceleration
of the relevant electrons and therefore implicit
in the derivation of the formula (\ref{omegacjack})
is a subtle role for dynamics~\cite{newlimit}.
From a field-theory perspective the process of
synchrotron-radiation emission
can be described in terms
of Compton scattering of the electrons
with the virtual photons of the magnetic field.
One would therefore
be looking deep into the dynamical features of the theory.

The minimal AEMNS test theory does assume a
modified dispersion relation
of the type (\ref{displeadbis}) universally applied
to all particles,
but it is a pure-kinematics framework and, since the analysis
crucially involves some aspects of dynamics,
it cannot be tested using a Crab-nebula
synchrotron-radiation analysis.

The GPMP test theory relies on a description of dynamics
within the framework of effective low-energy theory,
but, as mentioned, this in turn ends up implying that it is not
possible to assume
that a dispersion relation
of the type (\ref{displeadbis}) universally applies
to all particles.
Actually the two polarizations of photons must, within
this framework, satisfy different (opposite-sign Planck-scale
corrections) dispersion relations. And for the description
of electrons one naturally encounters at least two more free parameters.
The only constraint that one could conceivably obtain for the GPMP
test theory from the Crab-nebula
synchrotron-radiation analysis would simply exclude\footnote{Even
the possibility to derive any sort of
constraint on the electron-dispersion-relation parameters
is not guaranteed. In fact, as observed in the latest version
of Ref.~\cite{tedsteck}, one might be unable to exclude the
possibility that
the Crab-nebula synchrotron radiation be due to positron
(rather than electron) acceleration.}
that both
the electron-dispersion-relation parameters be negative
({\it i.e.} exclude
that both helicities of the electron would be characterized
by a dispersion relation
of the type (\ref{displeadbis}) with negative $\eta$ and $n=1$).

In particular, the case
which I characterized
as the ``minimal GPMP test theory",
where the two helicities of the electrons
carry opposite-sign modifications of the dispersion relation,
would automatically evade this type of constraint
from the Crab-nebula
synchrotron-radiation analysis (since the two helicities are
affected by opposite-sign modifications of the dispersion relation,
at least one of them must be a positive-sign-type modification).

\section{Summary and outlook}
\noindent
Quantum-Gravity Phenomenology
has already reached its first goal:
a sizeable community now works on the quantum-gravity problem
with the awareness that there is a chance to test
(at least some) Planck-scale effects.
In reaching this first goal it was sufficient (and even,
in a certain sense, necessary)
to proceed with simple intuitive arguments, but
the further development of quantum-gravity phenomenology
requires us to adopt the standards of other branches of phenomenology,
such as particle-physics phenomenology.
In particular, the progress of experimental
limits must be charted in terms of commonly-adopted,
and carefully crafted, test theories of the new
Planck-scale effects.

The fact that some Planck-scale pictures of spacetime physics
are falsifiable is more and more robustly established, but
in many cases we only see a path toward falsifiability rather
having achieved already the results needed for
a ``critical test of a theory'' (a test that could be used, in case of
contrary experimental results, to discard the
relevant Planck-scale picture of spacetime physics).
This point of the falsifiability of some relevant theories
is crucial for establishing quantum-gravity research as a truly
scientific endeavor.
The proposal of test theories must of course reflect the status
of our analysis of the falsifiability of quantum-spacetime/quantum-gravity
theories. In an appropriate sense the test theories must bridge the gap
between quantum-gravity theories and experiments.
They must be such that the experimental limits on the parameters of
the test theories will naturally translate into direct limits
on some relevant quantum-gravity theories, as soon as some
falsifiable features of the quantum-gravity theory are fully established.

It is of course meaningless to compare limits obtained
within different test theories. And there is no scientific content
in an experimental limit claimed on a vaguely defined test theory.
For example,
in the recent literature there has been a proliferation of
papers claiming to improve limits on Planck-scale modifications
of the dispersion relation, but the different studies were simply
considering the same type of dispersion relation within
significantly different test theories.
These results, which were presented as a gradual improvement in the
experimental limits on Planck-scale modifications of the dispersion relation,
were actually only a series of papers proposing more and more
(some better, some worse)
different examples of test theories in which a Planck-scale modification
of the dispersion relation can be accommodated.
Each paper was proposing a {\underline{different}} test theory and
deriving limits on that specific test theory.

In order to illustrate these issues in the context of
a specific example of quantum-gravity-phenomenology work,
in the second part of these lectures I focused on the example
of the phenomenology of Planck-scale modifications of
the dispersion relation.
I considered two examples of test theories,
the AEMNS test theory and the GPMP test theory.
These two test theories, although usually not explicitly fully
characterized in the relevant papers, are among the most
studied in the case of
Planck-scale modifications of
the dispersion relation.

I also stressed  that a phenomenology should build
its strength gradually. Within a given set of hypothesis
one first sets up a reduced parameter space, and only once
that reduced parameter space is ruled out by data one considers
the possibility of wider parameter spaces.
In the context here of interest the minimal AEMNS test theory,
described in Subsection~4.6,
and the minimal GPMP test theory,
described in Subsection~4.8,
appear to provide valuable starting points.

In particular, these two test theories can be representative
of two types of attitudes that are emerging in the
quantum-gravity-phenomenology community
concerning the possibility of describing dynamical effects
within the framework of effective low-energy field theory.
The fact that both in the study of noncommutative spacetimes
and in the study of Loop Quantum Gravity, the two quantum
pictures of spacetime that provide the key sources of motivation
for research on Planck-scale modifications of the dispersion relation,
we are really only starting to understand some aspects of kinematics,
but we are still missing any robust result on dynamics,
encourages an approach to phenomenology which
is correspondingly prudent with respect to the description
of dynamics.
The phenomenologist is therefore confronted with two options:
For those who are most concerned about the status of the
description of dynamics, the pure-kinematics
minimal AEMNS test theory provides a rather reasonable starting point
for phenomenology work.
For those who are willing to set aside these concerns,
and go ahead with the effective-field-theory description,
the minimal GPMP test theory could provide a valuable starting
point. It is interesting that, while the phenomenology based
on pure kinematics is allowed to start with the assumption
of full universality of the modification of the dispersion relation,
the choice of describing dynamics in terms of an effective low-energy
field theory forces upon us from the very beginning a nonuniversality
of the effects, with the correlation between polarization
and sign of the modification for photons
(and, with the additional natural assumption of no net effect on
randomly composed beams, one then can introduce for fermions
ana analogous correlation between helicity
and sign of the modification).
This plays a key role in the phenomenology.

In the Subsections~4.9,4.10,4.11,4.12
I have considered a few examples of phenomenological analyses
which exposed very clearly the type of
differences that one can encounter comparing
the indications of preliminary
sensitivity estimates and the outcome of more robust analyses
supported by test theories.
The time-of-travel analyses described in Subsection~4.9
can be used to constrain the photon dispersion relation
both in the AEMNS and in the GPMP test theory, but the strategy
may be somewhat different: while in the AEMNS test theory
one can only exploit the energy dependence of the new effects,
in the GPMP test theory the additional polarization dependence
can also be exploited.
The type of analysis of the cosmic-ray spectrum
described in Subsection~4.10 is also applicable to both
test theories, but also in that case some differences must be taken
into account.
In particular, by obtaining good-quality data on the cosmic-ray spectrum
around the GZK scale we might be in a position to
completely rule out the minimal GPMP test theory,
and to rule out the negative-$\eta$
case for the minimal AEMNS test theory.
The photon-stability analysis described in Subsection~4.11,
which received much attention in the literature, actually
turned out to be affected by severe limitations
in constraining the parameter spaces
of the  minimal AEMNS
and the minimal GPMP test theories: photon-stability analyses must
be treated prudently from a AEMNS perspective
because in principle kinematics is insufficient for establishing
the probability of particle decay (whereas kinematics is enough
for establishing stability), and photon-stability analyses only
lead to rather weak limits on the minimal-GPMP parameter space
because of the polarization dependence expected in that test theory
(one would need an ideally polarized beam of ultra-high-energy photons
in order to be able to infer some constraint on the GPMP test theory).
The Crab-nebula
synchrotron-radiation analysis, whose preliminary analysis
had also raised high hopes, when set up within the test theories
here of interest also proves to be largely ineffective:
it is not applicable to the AEMNS test theory (once again because
of the role that some aspects of dynamics play in the analysis)
and it also leads to no constraint on the minimal GPMP test theory.

While, consistently with the objectives of these lectures,
it was for me sufficient here to discuss this comparison of
test theories to data at a semi-quantitative level, the striking
results of this comparison, showing that the analysis at the test-theory
level can have very different outcome with respect to the usual
preliminary sensitivity estimates,
should provide motivation for
future publications with detailed quantitative analyses
of the emerging experimental bounds.


\begin{thebibliography}{99.}
%
%
%




\bibitem{stachelearly} J.~Stachel: Early History of
Quantum Gravity. In \textit{Black Holes,
Gravitational Radiation and the Universe},
ed by
B.R.~Iyer, B. Bhawal eds.~(Kluwer Academic Publisher,
Netherlands, 1999).

\bibitem{string1} M.B.~Green, J.H.~Schwarz and E.~Witten:
\textit{Superstring theory}
(Cambridge Univ. Press, Cambridge, 1987)

\bibitem{string2} J.~Polchinski: \textit{Superstring
Theory and Beyond},
(Cambridge University Press, Cambridge, 1998).

\bibitem{crLIVING} C.~Rovelli: gr-qc/9710008,
{Living Rev.~Rel.}~\textbf{1}, 1 (1998).

\bibitem{ashtNEW} A.~Ashtekar:
gr-qc/0112038.

\bibitem{leeLQGrev} L.~Smolin:
hep-th/0303185.

\bibitem{thieREV} T.~Thiemann: gr-qc/0210094,
Lect.~Notes Phys.~\textbf{631}, 41 (2003).

\bibitem{connesbook} A.~Connes:
\textit{Noncommutative Geometry},
(Academic Press, 1995).

\bibitem{majidbok} S.~Majid:
\textit{Foundations of Quantum Group Theory},
(Cambridge Univ Pr, 1995).

\bibitem{polonpap} G. Amelino-Camelia:
gr-qc/9910089, {Lect.~Notes Phys.}~\textbf{541}, 1 (2000).

\bibitem{iqgr} D.V.~Ahluwalia:
gr-qc/0202098.

\bibitem{cow} R.~Colella, A.W.~Overhauser and S.A.~Werner:
Phys. Rev. Lett.~\textbf{34}, 1472 (1975).

\bibitem{dharamCOW1} D.V.~Ahluwalia:
gr-qc/9903074, {\ Nature} \textbf{398}, 199 (1999).

\bibitem{sakurai} J.J.~Sakurai:
\textit{ Modern Quantum Mechanics}
(Addison Wesley, New York 1993).

\bibitem{gasperiniEP} M.~Gasperini:
{\ Phys.~Rev.}~D \textbf{38}, 2635 (1988).

\bibitem{dharamEP} G.Z.~Adunas, E.~Rodriguez-Milla and D.V.~Ahluwalia:
{\ Gen.~Rel.~Grav.}~\textbf{33}, 183 (2001).

\bibitem{cowEPviol} K.C.~Littrel, B.E.~Allman and S.A.~Werner:
{\ Phys.~Rev.}~A \textbf{56}, 1767 (1997).

\bibitem{dharamCOW2} D.V.~Ahluwalia:
gr-qc/0009033.

\bibitem{anan1} J.~Anandan: Phys.~Lett.~A \textbf{105}, 280 (1984);
{\ Class.~Quant.~Grav.}~\textbf{1}, 151 (1984).

\bibitem{anan2} A.K.~Jain et al:
Phys.~Rev.~Lett.~\textbf{58}, 1165 (1987).

\bibitem{led}
N.~Arkani-Hamed, S.~Dimopoulos and G.~Dvali:
{\ Phys.~Lett.}~B \textbf{429}, 263 (1998).

\bibitem{chrisreview}) C.J. Isham:
\textit{Structural issues in quantum gravity}.
In \textit{Proceedings of General relativity and gravitation 1995}
(World Scientic, Singapore 1997).


\bibitem{ehns} J.~Ellis, J.S.~Hagelin, D.V.~Nanopoulos and M.~Srednicki:
{\ Nucl.~Phys.}~B \textbf{241}, 381 (1984).

\bibitem{huetpesk}
P.~Huet and M.E.~Peskin:
{\ Nucl.~Phys.}~B \textbf{434}, 3 (1995).

\bibitem{emln} J.~Ellis, J.~Lopez, N.E.~Mavromatos and D.V.~Nanopoulos:
{\ Phys.~Rev.}~D \textbf{53}, 3846 (1996).

\bibitem{floreacpt} F.~Benatti and R.~Floreanini:
{\ Nucl.~Phys.}~B \textbf{488}, 335 (1997).

\bibitem{emn} J.~Ellis, N.~Mavromatos and D.V.~Nanopoulos:
Phys.~Lett.~B \textbf{293}, 37 (1992);
E.~Gravanis and N.E.~Mavromatos: hep-th/0108008.

\bibitem{kostcpt} V.A.~Kostelecky and R.~Potting:
{\ Phys.~Rev.}~D \textbf{51}, 3923 (1995);
O.~Bertolami, D.~Colladay, V.A.~Kostelecky and R.~Potting:
{\ Phys.Lett.}~B \textbf{395}, 178 (1997).

\bibitem{kostrenorm} D.~Colladay and V.A.~Kostelecky, hep-ph/9809521,
Phys.~Rev.~D \textbf{58}, 116002 (1998).

\bibitem{perci} I.C.~Percival:
Physics World \textbf{10} (No 3), 43 (1997).

\bibitem{percistru} I.C.~Percival and W.T.~Strunz:
{\ Proc.~Roy.~Soc.}~A \textbf{453}, 431 (1997).

\bibitem{web1} A.~Apostolakis et al:
hep-ex/9903005,
Phys.~Lett.~B \textbf{452}, 425 (1999).

\bibitem{web2} J.M.~Link et al:
hep-ex/0208034,
Phys.~Lett.~B \textbf{556}, 7 (2003).

\bibitem{ashtereview} A.~Ashtekar:
gr-qc/9901023.

\bibitem{crHISTO} C.~Rovelli:
gr-qc/0006061.

\bibitem{leePW} L.~Smolin:
{\ Physics World} \textbf{12}, 79 (1999).

\bibitem{carlip} S.~Carlip:
gr-qc/0108040,
{\ Rept.~Prog.~Phys.}~\textbf{64}, 885 (2001).

\bibitem{grbgac} G.~Amelino-Camelia, J.~Ellis, N.E.~Mavromatos,
D.V.~Nanopoulos and S.~Sarkar:
astro-ph/9712103,
{\ Nature} \textbf{393}, 763 (1998).

\bibitem{billetal} S.D.~Biller et al:
{\ Phys.~Rev.~Lett.}~\textbf{83}, 2108 (1999).

\bibitem{gacgwi} G. Amelino-Camelia:
gr-qc/9808029,
{\ Nature} \textbf{398}, 216 (1999);
gr-qc/0104086, {\ Nature} \textbf{410}, 1065 (2001).

\bibitem{bignapap} G. Amelino-Camelia:
gr-qc/9903080, {\ Phys.~Rev.}~D \textbf{62}, 024015 (2000).

\bibitem{nggwi} Y.J.~Ng and H.~van Dam:
gr-qc/9906003, Found.~Phys.~\textbf{30}, 795 (2000).

\bibitem{bignapatwo} G.~Amelino-Camelia:
gr-qc/0104005.

\bibitem{gaclaem} G.~Amelino-Camelia and
C.~Lammerzahl: gr-qc/0306019,
Class. Quant. Grav.~\textbf{21}, 899 (2004).

\bibitem{kifu} T.~Kifune:
{\ Astrophys.~J.~Lett.}~\textbf{518}, L21 (1999).

\bibitem{ita} R.~Aloisio, P.~Blasi, P.L.~Ghia and A.F.~Grillo:
Phys.~Rev.~D \textbf{62}, 053010 (2000).

\bibitem{aus} R.J.~Protheroe and H.~Meyer:
{\ Phys.~Lett.}~B \textbf{493}, 1 (2000).

\bibitem{gactp}  G. Amelino-Camelia and T. Piran:
{\ Phys.~Rev.}~D \textbf{64}, 036005 (2001).

\bibitem{jaco} T.~Jacobson, S.~Liberati and D.~Mattingly:
hep-ph/0112207, Phys.~Rev.~D \textbf{66},  081302 (2002).

\bibitem{gacpion} G.~Amelino-Camelia: gr-qc/0107086,
{\ Phys.~Lett.}~B \textbf{528}, 181 (2002).

\bibitem{orfeupion} O.~Bertolami:
hep-ph/0301191.

\bibitem{hooftlorentz} G.~`t Hooft:
{\ Class.~Quant.~Grav.}~\textbf{13}, 1023 (1996).

\bibitem{gacdsr} G.~Amelino-Camelia: gr-qc/0012051,
{\ Int.~J.~Mod.~Phys.}~D \textbf{11}, 35 (2002);
hep-th/0012238,
{\ Phys.~Lett.}~B \textbf{510}, 255 (2001).

\bibitem{majrue} S.~Majid and H.~Ruegg:
{\ Phys.~Lett.}~B \textbf{334}, 348 (1994).

\bibitem{kpoinap} J.~Lukierski, H.~Ruegg and W.J.~Zakrzewski:
{\ Ann.~Phys.}~\textbf{243}, 90 (1995).

\bibitem{gampul} R.~Gambini and J.~Pullin:
{\ Phys.~Rev.}~D \textbf{59}, 124021 (1999).

\bibitem{mexweave} J.~Alfaro, H.A.~Morales-Tecotl and L.F.~Urrutia:
{\ Phys.~Rev.~Lett.}~\textbf{84}, 2318 (2000).

\bibitem{gacmajid} G.~Amelino-Camelia and  S.~Majid:
{\ Int.~J.~Mod.~Phys.}~A \textbf{15}, 4301 (2000).

\bibitem{ahlucpt} D.V.~Ahluwalia:
{\ Mod.~Phys.~Lett.}~A \textbf{13}, 2249 (1998).

\bibitem{muracpt} H.~Murayama and T.~Yanagida:
{\ Phys.~Lett.}~B \textbf{520}, 263 (2001).

\bibitem{nickcpt} N.E.~Mavromatos:
hep-ph/0402005.

\bibitem{garaytest} L.J.~Garay:
{\ Phys.~Rev.~Lett.}~\textbf{80}, 2508 (1998).

\bibitem{gacpi} G.~Amelino-Camelia and S.-Y.~Pi:
hep-ph/9211211,
Phys.~Rev.~D \textbf{47}, 2356 (1993).

\bibitem{susskind} A.~Matusis, L.~Susskind and N.~Toumbas:
{\ JHEP} \textbf{0012}, 002 (2000).

\bibitem{dougnekr} N.R.~Douglas and N.A.~Nekrasov:
Rev.~Mod.~Phys.~\textbf{73}, 977 (2001).

\bibitem{gianlucaken} G.~Amelino-Camelia, G.~Mandanici and K.~Yoshida:
hep-th/0209254,
JHEP \textbf{0401}, 037 (2004).

\bibitem{skepticNCFT} P.~Fischer and V.~Putz:
hep-th/0306099,
Eur.~Phys.~J.~C \textbf{32}, 269 (2004)

\bibitem{wessLANGUAGE} J.~Madore, S.~Schraml, P.~Schupp and J.~Wess:
hep-th/0001203,
Eur.~Phys.~J.~C \textbf{16}, 161 (2000).

\bibitem{dopl1994} S.~Doplicher, K.~Fredenhagen and J.E.~Roberts:
Phys.~Lett.~B \textbf{331}, 39 (1994).

\bibitem{dsrmost} J.~Kowalski-Glikman:
hep-th/0102098,
Phys.~Lett.~A \textbf{286}, 391 (2001);
S.~Alexander and J.~Magueijo:
hep-th/0104093;
R.~Bruno, G.~Amelino-Camelia and J.~Kowalski-Glikman:
hep-th/0107039,
Phys.~Lett.~B \textbf{522}, 133 (2001);
J.~Magueijo and L.~Smolin:
gr-qc/0207085, Phys.~Rev.~D \textbf{67}, 044017 (2003);
J.~Kowalski-Glikman and S.~Nowak:
hep-th/0304101,
Class.~Quant.~Grav.~\textbf{20}, 4799 (2003).

\bibitem{lukieFT} P.~Kosinski, J.~Lukierski and P.~Maslanka:
Czech.~J.~Phys.~\textbf{50}, 1283 (2000).

\bibitem{gacmich} G.~Amelino-Camelia and M.~Arzano:
hep-th/0105120,
Phys.~Rev.~D \textbf{65}, 084044 (2002);
A.~Agostini, G.~Amelino-Camelia, F.~D'Andrea:
hep-th/0306013.

\bibitem{wesskappa} M.~Dimitrijevic, L.~Jonke, L.~Moller,
E.~Tsouchnika , J.~Wess and M.~Wohlgenannt:
hep-th/0307149,
Eur.~Phys.~J.~C \textbf{31}, 129 (2003).

\bibitem{simonecarlo} C.~Rovelli and S.~Speziale:
gr-qc/0205108.

\bibitem{areanew} G.~Amelino-Camelia:
gr-qc/0205125.

\bibitem{kodadsr} G.~Amelino-Camelia, L.~Smolin and A.~Starodubtsev:
hep-th/0306134,
Class.~Quant.~Grav.~\textbf{21}, 3095 (2004).

\bibitem{jurekkodadsr} L.~Freidel, J.~Kowalski-Glikman and L.~Smolin:
hep-th/0307085.

\bibitem{gampulDENSI} R.~Gambini, R.~Porto and J.~Pullin:
gr-qc/0305098.

\bibitem{Kosinski:2002gu} P.~Kosinski and P.~Maslanka: hep-th/0211057,
Phys.~Rev.~D \textbf{68}, 067702 (2003).

\bibitem{Mignemi:2003ab} S.~Mignemi: hep-th/0302065,
Phys.~Lett.~A \textbf{316}, 173 (2003).

\bibitem{Daszkiewicz:2003yr} M.~Daszkiewicz, K.~Imilkowska
and J.~Kowalski-Glikman:
hep-th/0304027.

\bibitem{jurekREV} J.~Kowalski-Glikman:
hep-th/0312140.

\bibitem{aemn1} G.~Amelino-Camelia, J.~Ellis, N.E.~Mavromatos
and D.V.~Nanopoulos:
hep-th/9605211,
Int.~J.~Mod.~Phys.~A \textbf{12}, 607 (1997).

\bibitem{gianluFranc} G.~Amelino-Camelia, F.~D'Andrea
and G.~Mandanici:
hep-th/0211022, JCAP \textbf{0309}, 006 (2003).

\bibitem{nycksync} J.~Ellis, N.E.~Mavromatos and A.S.~Sakharov:
astro-ph/0308403.

\bibitem{rob} R.C.~Myers and M.~Pospelov:
hep-ph/0301124, Phys.~Rev.~Lett.~\textbf{90}, 211601 (2003).

\bibitem{suda1} A.~Perez and D.~Sudarsky: gr-qc/0306113.

\bibitem{suda2} J.~Collins, A.~Perez, D.~Sudarsky, L.~Urrutia and H.~Vucetich:
gr-qc/0403053.

\bibitem{mg10qg1}  G.~Amelino-Camelia:
gr-qc/0402009.

\bibitem{glast}  J.P.~Norris, J.T.~Bonnell, G.F.~Marani and
J.D.~Scargle:
astro-ph/9912136;
A.~de Angelis:
astro-ph/0009271.

\bibitem{piranKARP} T.~Piran,
astro-ph/0407462.

\bibitem{grbNEUTRINOnew} P.~Meszaros, S.~Kobayashi, S.~Razzaque, B.~Zhang:
astro-ph/0305066.

\bibitem{dsrphen} G.~Amelino-Camelia, J.~Kowalski-Glikman,
G.~Mandanici and A.~Procaccini:
gr-qc/0312124.

\bibitem{nguhecr} Y.J.~Ng, D.S.~Lee, M.C.~Oh, and H.~van Dam:
Phys.~Lett.~B \textbf{507}, 236 (2001);
G. Amelino-Camelia, Y.J. Ng, and H. van Dam: gr-qc/0204077,
Astropart.~Phys.~\textbf{19}, 729 (2003).

\bibitem{agasa} M.~Takeda et al:
Phys.~Rev.~Lett.~\textbf{81}, 1163 (1998).

\bibitem{auger} J.~Blumer:
J.~Phys.~G \textbf{29}, 867 (2003).

\bibitem{seth} T.J.~Konopka and S.A.~Major:
New J.~Phys.~\textbf{4}, 57 (2002);
D.~Heyman, F.~Hinterleitner, and S.~Major:
gr-qc/0312089, Phys.~Rev.~D~\textbf{69}, 105016 (2004).

\bibitem{jacoNATv1} T.~Jacobson, S.~Liberati and D.~Mattingly:
arXiv.org/abs/astro-ph/0212190v1

\bibitem{newlimit} G.~Amelino-Camelia:
gr-qc/0212002.

\bibitem{jaconature} T.~Jacobson, S.~Liberati and D.~Mattingly:
arXiv.org/abs/astro-ph/0212190v2,
Nature \textbf{424}, 1019 (2003).

\bibitem{tedreply} T.~Jacobson, S.~Liberati and D.~Mattingly:
gr-qc/0303001.

\bibitem{carrosync} S.~Carroll: Nature {\bf 424}, 1007 (2003).

\bibitem{tedsteck} T.A.~Jacobson, S.~Liberati, D.~Mattingly and F.W.~Stecker:
astro-ph/0309681.

\bibitem{jackson} J.D.~Jackson, \textit{Classical Electrodynamics},
3rd edn
(J.~Wiley \& Sons, New York 1999).

\end{thebibliography}
\end{document}